\documentclass[12pt]{iopart}

\usepackage{lineno,hyperref}
\usepackage{graphics}
\usepackage{subcaption}
\expandafter\let\csname equation*\endcsname\relax
\expandafter\let\csname endequation*\endcsname\relax
\usepackage{amsmath}
\usepackage{bbold}
\usepackage{physics}
\usepackage{xcolor}
\usepackage{url}
\usepackage[export]{adjustbox}

\begin{document}

\title[Bethe-Salpeter equation for absorption and scattering spectroscopy]{Bethe-Salpeter equation for absorption and scattering spectroscopy: Implementation in the \texttt{exciting} code}

\author{Christian Vorwerk$^{1,2}$, Benjamin Aurich$^{1,2}$, Caterina Cocchi$^{1,2}$, and Claudia Draxl$^{1,2}$}
\address{$^1$ Institut f\"ur Physik and IRIS Adlershof,Humboldt-Universit\"at zu Berlin, Berlin, Germany}
\address{$^2$ European Theoretical Spectroscopic Facility (ETSF)}

\begin{abstract}
The Bethe-Salpeter equation for the electron-hole correlation function is the state-of-the-art formalism for optical and core spectroscopy in condensed matter. Solutions of this equation yield the full dielectric response, including both the absorption and the inelastic scattering spectra. Here, we present an efficient implementation within the all-electron full-potential code \texttt{exciting}, which employs the linearized augmented plane-wave (L)APW+LO basis set. Being an all-electron code, \texttt{exciting} allows the calculation of optical and core excitations on the same footing. The implementation fully includes the effects of finite momentum transfer which may occur in inelastic x-ray spectroscopy and electron energy-loss spectroscopy. Our implementation does not require the application of the Tamm-Dancoff approximation that is commonly employed in the determination of absorption spectra in condensed matter. The interface with parallel linear-algebra libraries enables the calculation for complex systems.
The capability of our implementation to compute, analyze, and interpret the results of different spectroscopic techniques is demonstrated by selected examples of prototypical inorganic and organic semiconductors and insulators.
\end{abstract}
\maketitle
\section{Introduction}
The Bethe-Salpeter equation (BSE) is the state-of-the-art method to describe light absorption in crystalline materials~\cite{hedi65pr,hybe-loui85prl,Strinati1988,onid+95prl,albr+97prb,bene+98prl,rohl-loui98prl} and molecular systems in their condensed phase~\cite{gross+01prl,pusc-ambr02prl,humm+04prl,humm-ambr05prb,tiag-chel05ssc,hahn+05prl,palu+09jcp,fabe+14ptrsa,cocc-drax15prb,hiro+15prb,brun+15jcp,hung+16prb}. This approach enables the calculation of the dynamical polarizability including the effects of the electron-hole interaction and thus yields insight into energy, strength, and character of (bound) excitonic states. The application of the BSE formalism to treat transitions from core electrons also gives access to x-ray absorption  and inelastic x-ray scattering spectra~\cite{vinson+11PRB,vins-rehr12prb,nogu+15jctc,gilm+15cpc,cocc+16prb,foss+17prb,lask-blah10prb,vorw+17prb,drax-cocc17condmat,olov+09prb,olov+09jpcm,olov+11prb,olov+13jpcm,cocc-drax15prb1,vorw+18jpcl}. 

BSE implementations typically focus on the optical limit, where the momentum transferred from the photons to the electronic system can be neglected. While this approximation is justified for optical absorption spectra, in both inelastic x-ray scattering (IXS) and electron-energy loss spectroscopy (EELS), the influence of the photon momentum loss needs to be accounted for. BSE calculations at finite momentum transfer have been performed only in a limited number of works~\cite{gatt-sott13prb,cuda+13prb,fuga+15prb,cuda+16prl,kosk+17prb,spon+18prb}, which nonetheless demonstrate their relevance for accessing and complementing scattering spectroscopy experiments.

Furthermore, currently available BSE implementations for solids are typically limited to the Tamm-Dancoff approximation (TDA), where the coupling between excitations and de-excitations is neglected. While the TDA is reasonable for calculations related to optical absorption measurements of conventional inorganic semiconductors, the spectroscopic characterization of other material classes requires an extension of this approach, as, for instance, discussed in the context of molecular systems~\cite{grue+09nl,pusc+13condmat,ljun+15prb,rang+17jcp} and nanostructures~\cite{grue+09nl,rocc+14jctc}. 

All-electron full-potential methods, in particular those employing the LAPW+LO basis set, treat core and valence electrons on the same footing, thus enabling a reliable access to both valence and core excitations \cite{lask-blah10prb,vorw+17prb,drax-cocc17condmat,olov+09prb,olov+09jpcm,olov+11prb,olov+13jpcm,cocc-drax15prb1,vorw+18jpcl}. In this paper, we present the comprehensive implementation of the BSE formalism in the all-electron full-potential code \texttt{exciting}, with a focus on recent developments that comprise the construction and solution of the full BSE without the application of the Tamm-Dancoff approximation, the solution of the BSE beyond the optical limit, and the unified description of core and valence excitations. 
After reviewing the theoretical background for optical and core-level excitations, we describe the general structure of the code and address how we solve computational challenges within our implementation. We demonstrate the functionalities of the new developments with selected examples, ranging from optical to core spectroscopy and spanning a broad set of materials, including bulk semiconductors and insulators as well as organic crystalline structures.
\section{Theoretical Background}
\subsection{Dielectric linear response}
In linear response theory, both absorption and inelastic scattering spectra can be obtained from the macroscopic dielectric function $\epsilon_M(\mathbf{Q},\omega)$ of the system. This quantity is, in turn, connected to the \textit{microscopic} inverse dielectric function $\epsilon_{\mathbf{G}\mathbf{G}'}(\mathbf{q},\omega)$ by
\begin{equation}
\epsilon_M(\mathbf{Q}=\mathbf{q}+\mathbf{G},\omega)=\frac{1}{\epsilon^{-1}_{\mathbf{G},\mathbf{G}}(\mathbf{q},\omega)},
\end{equation}
where we have expressed the momentum-loss vector $\mathbf{Q}=\mathbf{G}+\mathbf{q}$ as the sum of the reciprocal lattice vector $\mathbf{G}$ and the vector $\mathbf{q}$ from the first Brillouin zone.
In linear response theory, the microscopic inverse dielectric function is obtained from the polarizability $P_{\mathbf{G}\mathbf{G}}(\mathbf{q},\omega)$ as \cite{Onida2002}
\begin{equation}
  \epsilon^{-1}_{\mathbf{G}\mathbf{G}'}(\mathbf{q},\omega) = 1 + v_{\mathbf{G}}(\mathbf{q}) P_{\mathbf{G}\mathbf{G}'}(\mathbf{q},\omega),
  \label{eqn:epsmofpbar}
\end{equation}
where $v$ is the bare Coulomb potential.
\subsection{The Bethe-Salpeter equation}
The polarizability $P(1,2)=P(\mathbf{r}_1,\mathbf{r}_2,t_1,t_2)$ of Eq.~\ref{eqn:epsmofpbar} is expressed within many-body perturbation theory (MBPT) in terms of one-particle Green's functions, $G_1$, and two-particle Green's functions, $G_2$. 
It is given by the following relation
\begin{equation}
P(1,2) = \mathrm{i} \left[ G_2(1,2,1^+,2^+) - G_1(1,1^+) G_1(2,2^+) \right],
\label{eqn:chiL}
\end{equation}
where we employ for the indexes the common shorthand notation $1 \equiv (\mathbf{r}_1, t_1)$ and $1^+ \equiv(\mathbf{r}_1,t_1+0^{+})$. 
Propagators are defined as
\begin{equation}
  G_1(1; 2) 
  = -\mathrm{i} \langle\hat{\mathcal{T}}\left(\hat{\mathrm{\Psi}}(1)\hat{\mathrm{\Psi}}^{\dagger}(2)\right)\rangle
  \label{eqn:G1}
\end{equation}
and
\begin{equation}
  G_2(1, 2; 1', 2') 
  = -\langle \hat{\mathcal{T}}\left(\hat{\mathrm{\Psi}}(1) \hat{\mathrm{\Psi}}(2)\hat{\mathrm{\Psi}}^{\dagger}(2') \hat{\mathrm{\Psi}}^{\dagger}\right)(1')\rangle,
  \label{eqn:G2}
\end{equation}
using the time-ordering operator $\hat{\mathcal{T}}$, and the Heisenberg field operators $\hat{\mathrm{\Psi}}$, $\hat{\mathrm{\Psi}}^{\dagger}$. Introducing the \textit{electron-hole correlation function} $L$ in the notation of Refs.~\cite{Strinati1988,Sagmeister,Onida2002} 
\begin{equation}
  L(1,2,1',2') = -G_2(1,2,1',2') + G_1(1,1') G_1(2,2'),
  \label{eqn:L}
\end{equation}
we obtain
\begin{equation}
  P(1,2) = -\mathrm{i} L(1,2,1^+,2^+).
  \label{eqn:chiL2}
\end{equation}
In order to calculate the dielectric function, we have to solve the BSE~\cite{Strinati1988} for $L$:
%
\begin{equation}
  L(1,2;1',2') = L_0(1,2;1',2') +  \int \mathbf{d}(3,4,5,6)\times L_0(1,4;1',3)\; \Xi(3,5;4,6)\; L(6,2;5,2'),
  \label{eqn:BSEinL}
\end{equation}
where $L_0(1,2;1',2') = G_1(1,2')G_1(2,1')$ describes the propagation of two independent particles, and $\Xi$ is the kernel accounting for the two-particle interactions. 
 The interaction kernel is given on the $GW$ level by~\cite{Onida2002,Strinati1988}
\begin{equation} \label{eqn:Xistatic1}
    \Xi(3,5;4,6) \approx - \mathrm{i} \delta(3,4)\delta(5,6)v(3,6)\delta(t_3-t_6)+ \mathrm{i} \delta(3,6) \delta(4,5) w(\mathbf{r}_3,\mathbf{r}_5) \delta(t_3 - t_5),
\end{equation}
where the first term describes the exchange interaction through the bare Coulomb potential $v$, while the second one accounts for the screened electron-hole attraction $w$. 
The time restrictions and the approximation of $\Xi$ entail that the resulting BSE only depends on one time difference. 

\subsection{BSE in matrix form}
Since the quantities $L(1,2;1',2')$ and $L_0(1,2;1',2')$ in the BSE (Eq.~\ref{eqn:BSEinL}) depend on four points in space and time, they can be represented as matrices in the basis formed by products of single-particle wavefunctions $\phi_{i \mathbf{k}}$. These products form the \textit{transition space} in the independent particle picture, where the single-particle wavefunctions represent the initial and final state of the transition. Typically, this basis is split into the \textit{resonant} part, \textit{i.e.} transitions from occupied to unoccupied states with positive transition energies, and the \textit{anti-resonant} part, \textit{i.e.} transitions from unoccupied to occupied states with negative transition energy. Here, we define such a basis with the functions $\Upsilon^\mathrm{r}$ and $\Upsilon^\mathrm{a}$~\cite{Sander2015} for the resonant and anti-resonant space, respectively:
\begin{equation}
    \Upsilon^\mathrm{r}_{\alpha,\mathbf{q}}(\mathbf{r},\mathbf{r}') = 
    \phi_{o \mathbf{k}_+}(\mathbf{r})
    \phi^*_{u \mathbf{k}_-}(\mathbf{r}') 
    \label{eqn:2particlebasis_qsym_res}
\end{equation}
and
\begin{equation}
    \Upsilon^{\mathrm{a}}_{\alpha,\mathbf{q}}(\mathbf{r},\mathbf{r}') =
    \phi_{u (-\mathbf{k}_-)}(\mathbf{r})\phi^*_{o (-\mathbf{k}_+)}(\mathbf{r}'),
  \label{eqn:2particlebasis_qsym_ares}
  \end{equation}
with the index $o$ ($u$) denoting occupied (unoccupied) states, and the $\mathbf{k}$-point set chosen such that $\mathbf{k}_{\pm}=\mathbf{k}\pm \frac{\mathbf{q}}{2}$. $\alpha$ is a combined index $\alpha \leftrightarrow \left\{ o, u, \mathbf{k} \right\}$ which, together with the index $\mathbf{q}$, uniquely labels independent particle transitions from $\phi_{o\mathbf{k}_-}$ to $\phi_{u\mathbf{k}_+}$. This specific choice of basis functions allows us to exploit the symmetry properties $\phi_{n \mathbf{k}}(\mathbf{r})=\phi^{*}_{n -\mathbf{k}}(\mathbf{r})$ and $\epsilon_{n \mathbf{k}}=\epsilon_{n -\mathbf{k}}$ of the Bloch states under time-reversal, such that \cite{Sander2015,aurich-thesis2017}
\begin{equation}
  \Upsilon^\mathrm{a}_{\alpha,\mathbf{q}}(\mathbf{r},\mathbf{r}') = \Upsilon^\mathrm{r}_{\alpha,\mathbf{q}}(\mathbf{r}',\mathbf{r}).
  \label{eqn:2particle_sym}
\end{equation}
The matrix elements of $L$ in this basis are obtained as
\begin{equation}
  L_{ij}(\mathbf{q}) = \int d^3{r_1} d^3{r_1'} d^3{r_2} d^3{r_2'}\times \Upsilon^*_{i\mathbf{q}}(\mathbf{r}_1, \mathbf{r}_1') L(\mathbf{r}_1,\mathbf{r}_2,\mathbf{r}_1',\mathbf{r}_2')\Upsilon_{j\mathbf{q}}(\mathbf{r}_2', \mathbf{r}_2),
  \label{eqn:matrixelements}
\end{equation}
where $i$ and $j$ combine the indices of the transition ($\alpha$) and of the resonant or anti-resonant subspace (r or a). This choice of basis set has the additional advantage that the independent-particle correlation function $L_0$ in Eq.~\ref{eqn:BSEinL} becomes diagonal, and the inverse $L_0^{-1}$ takes the form
\begin{equation}
  \mathrm{L}^{-1}_0(\mathbf{q}, \omega) =
  -\left[ 
  \mqty( \mathrm{E}^\text{ip}(\mathbf{q}) & 0 \\ 0 & \mathrm{E}^\mathrm{ip}(\mathbf{q}) )
  - \omega \mqty( \mathbb{1} & 0 \\ 0 & -\mathbb{1} ) \right],
  \label{eqn:chi0mat-tr}
\end{equation}
where 
\begin{equation}
  E^\mathrm{ip}_{\alpha,\alpha'}(\mathbf{q}) = 
  \left(\epsilon_{u, \mathbf{k}_-} - \epsilon_{o, \mathbf{k}_+}\right) \delta_{\alpha,\alpha'}
  \label{eqn:emat}
\end{equation}
contains the independent-particle transition energies. 

We now write the BSE (Eq.~\ref{eqn:BSEinL}) as a matrix equation in the basis of Eqs.~\ref{eqn:2particlebasis_qsym_res} and \ref{eqn:2particlebasis_qsym_ares}. For crystalline systems, the response function can be written as a sum of functions defined for each point in the Brillouin zone (BZ): $L = \sum_{\mathbf{q}} L_\mathbf{q}$. Thus, the BSE can be solved individually for each $\mathbf{q}$-point such that we obtain
\begin{equation}
  L(\mathbf{q}, \omega) = \left[ L^{-1}_0(\mathbf{q},\omega) -
 \Xi(\mathbf{q}) \right]^{-1}.
  \label{eqn:chimatrixq}
\end{equation}
Inserting the explicit form of $L_0$ (Eq.~\ref{eqn:chi0mat-tr}) into this equation, we arrive at
\begin{equation}
  L(\mathbf{q}, \omega) 
  =
  -
  \left[ \mathrm{H}(\mathbf{q}) - \omega \mathrm{\Delta}\right]^{-1},
  \label{eqn:chiofh}
\end{equation}
where $\mathrm{H}(\mathbf{q})$ includes all frequency-independent terms and $\mathrm{\Delta} = \left(\begin{array}{cc}\mathbb{1} & 0 \\ 0 & -\mathbb{1}\end{array} \right)$. The matrix $\mathrm{H}(\mathbf{q})$ represents an effective Hamiltonian, the \textit{BSE Hamiltonian}, the eigenstates of which are also eigenstates of $L$.
%
%
Using time-reversal symmetry~\cite{Sander2015} and making use of the symmetry property of Eq.~\ref{eqn:2particle_sym}, the Hamiltonian becomes hermitian and takes the form:
\begin{equation}
  \mathrm{H}(\mathbf{q})
  =
  \left(\begin{array}{cc} \mathrm{A}(\mathbf{q}) & \mathrm{B}(\mathbf{q}) \\  \mathrm{B}(\mathbf{q}) & \mathrm{A}(\mathbf{q})\end{array} \right),
  \label{eqn:Hetr}
\end{equation}
with the diagonal block expressed by
\begin{equation}
  \mathrm{A}(\mathbf{q}) 
  = 
  \mathrm{E}^\mathrm{ip}(\mathbf{q})
  + 2 \gamma_\mathrm{x} \mathrm{V}^\mathrm{rr}(\mathbf{q})
  - \gamma_\mathrm{c} \mathrm{W}^\mathrm{rr}(\mathbf{q})
  \label{eqn:diagonalblock}
\end{equation}
and the coupling block being
\begin{equation}
  \mathrm{B}(\mathbf{q}) =
  2 \gamma_\mathrm{x} \mathrm{V}^{\mathrm{rr}}(\mathbf{q}) 
  - \gamma_\mathrm{c} \mathrm{W}^{\mathrm{ra}}(\mathbf{q}).
  \label{eqn:couplingblock}
\end{equation}
We have introduced the factors $\gamma_{x}$ and $\gamma_\mathrm{c}$ to account for the spin degree of freedom (see also Ref.~\cite{Puschnig}). 
Spin-singlet excitations are obtained by setting $\gamma_\mathrm{x}=1$ and $\gamma_\mathrm{c}=1$, while spin-triplet excitations are calculated with $\gamma_\mathrm{x}=0$ and $\gamma_\mathrm{c}=1$.

\section{LAPW+LO Basis}
In this section, we introduce the computational steps that are needed to solve the BSE in the \texttt{exciting} code.
\texttt{exciting} employs the (L)APW+lo basis set in the Kohn-Sham equations to compute valence and conduction states. These states then enter the expressions of the matrix elements of the BSE Hamiltonian. In this basis, the unit cell is divided into non-overlapping muffin-tin (MT) spheres centered at the atomic positions and the \textit{interstitial} space between the spheres. Different functions are employed in the two regions in order to account for both the rapid variation of the Kohn-Sham wavefunctions close to the nuclei and the smoother behavior in the interstitial region. In the MT sphere surrounding an atom $\alpha$, the wavefunctions are expanded in atomic-like basis functions $u_{l}^{\alpha}(r)Y_{lm}(\hat{r})$, while plane waves $\mathrm{e}^{-\mathrm{i}(\mathbf{G}+\mathbf{k})\mathbf{r}}$ are used in the interstitial region. As such, the basis functions $\phi_{\mathbf{k}+\mathbf{G}}$ are expressed as
\begin{equation}
 \phi_{\mathbf{k}+\mathbf{G}}(\mathbf{r})=\left\{ \begin{array}{cc}\frac{1}{\sqrt{\Omega}}\mathrm{e}^{-\mathrm{i}(\mathbf{G}+\mathbf{k})\mathbf{r}} & \mathbf{r} \in \mathrm{I} \\ \sum_{lm, p}A^{\mathbf{k}+\mathbf{G}}_{lm}u^{\alpha}_{l,p}(r)Y_{lm}(\hat{r}) & \mathbf{r} \in MT \end{array} \right. .
\end{equation}
Here, $\Omega$ is the unit-cell volume and $A^{\mathbf{k}+\mathbf{G}}_{lm, p}$ are expansion coefficients that ensure that the basis functions are continuous at the boundaries of the MT spheres. The radial functions $u^{\alpha}_{l, p}(r)$ are obtained from the solutions of the radial Schr\"odinger equation using the spherically averaged Kohn-Sham potential, where the index $p$ denotes $p$-th derivative with respect to the energy, \textit{i.e.} $u^{\alpha}_{l, p}=\frac{\partial^{p} u^{\alpha}_{l}}{\partial \epsilon ^{p}}$ . In order to increase the variational degrees of freedom in the MT spheres, local orbitals (LOs) $\phi_{\nu}(\mathbf{r})$ are used to complement the basis. These basis function are expressed as
\begin{equation}
 \phi_{\nu}(\mathbf{r})=\begin{cases}0 & \mathbf{r} \in \mathrm{I} \\ \delta_{\alpha \alpha_{\nu}}\delta_{ll_{\nu}}\delta_{mm_{\nu}}\sum_{p}B_{\nu, p}u^{\alpha}_{l, p}(r)Y_{lm}(\hat{r})  & \mathbf{r} \in MT \end{cases}.
\end{equation}
The local orbitals vanish outside of the MT spheres and the coefficients $B_{\nu, p}$ ensure that they are continuous and smooth at the MT-sphere boundary. As the LOs are added for specific MT spheres and $(lm)$-channels, they allow for a systematic improvement of the basis. For a review on the family of (L)APW+lo basis sets, see Ref.~\cite{Gulans2014}.
The eigenstates $\psi_{i \mathbf{k}}$ of the Kohn-Sham Hamiltonian are expressed in the LAPW+LO basis as
\begin{equation}\label{eqn:lapwlo-basis}
 \psi_{i \mathbf{k}}(\mathbf{r})=\sum_{\mathbf{G}} C_{i (\mathbf{k}+\mathbf{G})} \phi_{\mathbf{k}+\mathbf{G}}(\mathbf{r})+C_{i \nu \mathbf{k}}\phi_{\nu}(\mathbf{r})
 =\left\{ \begin{array}{cc}\frac{1}{\sqrt{\Omega}}\sum_{\mathbf{G}}C_{i (\mathbf{k}+\mathbf{G})}\mathrm{e}^{-\mathrm{i}(\mathbf{G}+\mathbf{k})\mathbf{r}} & \mathbf{r} \in \mathrm{I}\\  \sum_{lm} u^{i \mathbf{k}}_{l}(r)Y_{lm}(\hat{r}) & \mathbf{r} \in \mathrm{MT}\end{array}\right.
\end{equation}
where the radial functions are defined as 
\begin{equation}
u^{i \mathbf{k}}_l=\sum_{p}\sum_{\mathbf{G}}C_{i (\mathbf{k}+\mathbf{G})}A^{\mathbf{k}+\mathbf{G}}_{lm, p}u_{l,p}^{\alpha}(r)+\sum_{\nu}C_{i \nu \mathbf{k}}B_{\nu, p}u^{\alpha}_{l, p}(r). 
\end{equation}
$C_{i (\mathbf{k}+\mathbf{G})}$ and $C_{i \nu \mathbf{k}}$ are the single-particle eigenstates, obtained from the diagonalization of the Kohn-Sham Hamiltonian.

While the expansion in this basis is convenient for the extended valence and conduction states, the highly localized core states require a different treatment. As spin-orbit coupling can play a dominant role for these states, they are obtained from the solution of the radial Dirac equation in the spherically symmetrized crystal potential for each atomic site. The spinor solutions $\psi_{\kappa, M}$ of these equations can be written as
\begin{equation}
\psi_{\kappa, M}(\mathbf{r})=\left( \begin{array}{c} u_{\kappa}(r) \Omega_{\kappa, M}(\hat{r}) \\ -\mathrm{i}v_{\kappa}(r) \Omega_{-\kappa, M}(\hat{r}) \end{array} \right),
\end{equation}
where we have introduced an unique index $\kappa$ for a core state $^{2S+1}L_J$:
\begin{equation}
\kappa=\begin{cases}-L-1 & \mathrm{for}\; J=L+\frac{1}{2} \\L & \mathrm{for}\; J=L-\frac{1}{2} \end{cases}.
\end{equation}
The spherical part of the core wavefunctions $\psi_{\kappa, M}$ is given by the spin spherical harmonics $\Omega_{L,S,J,M}(\hat{r})$, while the radial functions $u_{\kappa}(r)$ for the large component and $-\mathrm{i}v_{\kappa}(r)$ for the small component, respectively, are given by the coupled radial Dirac equations
\begin{equation}
\frac{\partial u_{\kappa}}{\partial r}=\frac{1}{c}\left( v_{eff}-\epsilon_{\kappa}\right)v_{\kappa}+\left( \frac{\kappa-1}{r}\right) u_{\kappa}
\end{equation}
\begin{equation}
\frac{\partial v_{\kappa}}{\partial r}= -\frac{\kappa+1}{r}v_{\kappa}+2c\left[ 1+\frac{1}{2c^2}\left( \epsilon_{\kappa}-v_{eff}\right) \right],
\end{equation}
where $v_{eff}$ is the spherically averaged effective Kohn-Sham potential. In the calculation of matrix elements between core states and conduction states, the small component is neglected, and we obtain the wavefunction $\psi^{\alpha}_{\kappa, M}$ at an atomic site $\alpha$:
\begin{equation}\label{eq:core}
\psi^{\alpha}_{\kappa, M}(\mathbf{r})=\begin{cases} u_{\kappa, \alpha}(r_{\alpha})\Omega_{\kappa, M}(\hat{r}_{\alpha}) & \mathrm{for} \; r_{\alpha} \le R_{MT} \\ 0 & \mathrm{else} \end{cases}.
\end{equation}
More details about the treatment of core states in the LAPW+LO basis can be found in Ref.~\cite{vorwerkPRB}.
\section{Implementation}
In this section, we present the implementation of the BSE formalism in \texttt{exciting}. A schematic workflow is shown in Fig.~\ref{fig:workflow}, and more details are provided in the Appendix. Momentum and plane-wave matrix elements are central quantities, and are discussed in detail in this section. 

\subsection{Momentum and plane-wave matrix elements}
The momentum matrix elements $P^j_{nm \mathbf{k}}=\langle n \mathbf{k} |-\mathrm{i}\nabla_j| m \mathbf{k} \rangle$ between conduction and valence states are expanded in the LAPW+LO basis
\begin{equation}
\begin{aligned}
 P^j_{nm \mathbf{k}}&=\sum_{\mathbf{G}\mathbf{G}'} C^*_{n (\mathbf{k}+\mathbf{G})}C_{m (\mathbf{k}+\mathbf{G})} P^j_{\mathbf{G}\mathbf{G}'\mathbf{k}}+\sum_{\mathbf{G}\nu} C^*_{n (\mathbf{k}+\mathbf{G})}C_{m \nu \mathbf{k}} P^j_{\mathbf{G}\nu \mathbf{k}}\\
 &+\sum_{\nu' \mathbf{G}'} C^*_{m \nu' \mathbf{k}}C_{n (\mathbf{k}+\mathbf{G})}P^j_{\nu' \mathbf{G}' \mathbf{k}}+\sum_{\nu' \nu} C^*_{m \nu' \mathbf{k}}C_{n \nu \mathbf{k}}P^j_{\nu' \nu \mathbf{k}},
\end{aligned}
\end{equation}
where $C_{i \mathbf{k}}$ and $C_{i \nu \mathbf{k}}$ are the coefficients of Eq.~\ref{eqn:lapwlo-basis}, and $P^j_{\mathbf{G}\mathbf{G}'\mathbf{k}}$, $P^j_{\mathbf{G}\nu \mathbf{k}}$, $P^j_{\nu' \mathbf{G}' \mathbf{k}}$, and $P^j_{\nu' \nu}$ are LAPW-LAPW, LAPW-LO, LO-LAPW, and LO-LO momentum matrix elements, respectively, which are defined as
\begin{equation}
 \begin{aligned}
  &P^j_{\mathbf{G}\mathbf{G}'\mathbf{k}}=\langle \phi_{\mathbf{k}+\mathbf{G}} |-\mathrm{i}\nabla_j| \phi_{\mathbf{k}+\mathbf{G}'} \rangle\\
  &P^j_{\mathbf{G}\nu \mathbf{k}}=\langle \phi_{\mathbf{k}+\mathbf{G}} |-\mathrm{i}\nabla_j| \phi_{\nu} \rangle\\
  &P^j_{\nu' \mathbf{G}' \mathbf{k}}=\langle \phi_{\nu'} |-\mathrm{i}\nabla_j| \phi_{\mathbf{k}+\mathbf{G}'} \rangle\\
  &P^j_{\nu' \nu }=\langle \phi_{\nu'} |-\mathrm{i}\nabla_j| \phi_{\nu} \rangle.
 \end{aligned}
\end{equation}
These matrix elements of the general form $P^j_{ab}$ can furthermore be decomposed into contributions from the MT spheres $P^{j,MT}_{ab}$ and from the interstitial region $P^{j,MT}_{ab}$, such that we can write all plane-wave matrix elements as $P^j_{ab}=P^{j,MT}_{ab}+P^{j,I}_{ab}$. Since the local orbitals vanish in the interstitial region, only $P^j_{\mathbf{G}\mathbf{G}'}$ have a non-vanishing contribution. The interstitial part of the matrix elements are calculated analytically, as the action of the nabla-operator on plane waves can be determined analytically. In the MT spheres, the action of the nabla-operator is expanded in terms of spherical harmonics, $\nabla_j\left[ u_{lp}^{\alpha}(r)Y_{lm}(\hat{r})\right]=\sum_{l'm'}u^{\alpha,j}_{lmp,l'm'}Y_{l'm'}(\hat{r}) $. This expansion allows for the analytic evaluation of the spherical integral, while the radial integration within the MT spheres is performed numerically on a grid. 

For the matrix elements between a core state $(\kappa,M)$ and a conduction state $i$ at $\mathbf{k}$, the interstitial contribution vanishes, and the MT contribution is given by
\begin{equation}\label{eq:core-p}
 P^j_{\kappa M,i \mathbf{k}}=-\mathrm{i} \sum_{lm} \int d\Omega \; \Omega^*_{\kappa,M}(\hat{r}_{\alpha})\int_{R^{\alpha}_{MT}}dr \; r^2 u^*_{\kappa,\alpha}(r_{\alpha})\nabla_j\left[ u^{i\mathbf{k}}_{l}(r_{\alpha})Y_{lm}(\hat{r}_{\alpha}) \right].
\end{equation}
Analogous to the MT contributions of the matrix elements between conduction and valence states, the spherical integration is performed analytically, while the radial integration is performed numerically on a grid.

Plane-wave matrix elements $M_{mn \mathbf{k}}(\mathbf{G},\mathbf{q})=\langle m \mathbf{k}|\mathrm{e}^{-\mathrm{i}(\mathbf{G}+\mathbf{q})\mathbf{r}}|n (\mathbf{k}+\mathbf{q})\rangle$ are calculated accordingly:
\begin{equation}
  \begin{aligned}
    M_{mn \mathbf{k}}(\mathbf{G},\mathbf{q})&=\sum_{\mathbf{G}\mathbf{G}'} C^*_{m (\mathbf{k}+\mathbf{G})}C_{n (\mathbf{k}+\mathbf{q}+\mathbf{G})} M_{\mathbf{G}\mathbf{G}'\mathbf{k}}(\mathbf{G},\mathbf{q})+\sum_{\mathbf{G}\nu'} C^*_{m (\mathbf{k}+\mathbf{G})}C_{n \nu' \mathbf{k}} M_{\mathbf{G}\nu' \mathbf{k}}(\mathbf{G},\mathbf{q})\\
    &+\sum_{\nu \mathbf{G}'} C^*_{m \nu' \mathbf{k}}C_{n (\mathbf{k}+\mathbf{q}+\mathbf{G})}M_{\nu \mathbf{G}' \mathbf{k}}(\mathbf{G},\mathbf{q})+\sum_{\nu \nu'} C^*_{m \nu' \mathbf{k}}C_{n \nu \mathbf{k}+\mathbf{q}}M_{\nu \nu' \mathbf{k}}(\mathbf{G},\mathbf{q}),
  \end{aligned}
\end{equation}
where the LAPW-LAPW, LAPW-LO, LO, LAPW-LO, and LO-LO plane-wave matrix elements are given by
\begin{equation}
 \begin{aligned}
  &M_{\mathbf{G}\mathbf{G}'\mathbf{k}}(\mathbf{G},\mathbf{q})=\langle \phi_{\mathbf{k}+\mathbf{G}} |\mathrm{e}^{-\mathrm{i}(\mathbf{G}+\mathbf{q})\mathbf{r}}| \phi_{\mathbf{k}+\mathbf{q}+\mathbf{G}'} \rangle\\
  &M_{\mathbf{G}\nu \mathbf{k}}(\mathbf{G},\mathbf{q})=\langle \phi_{\mathbf{k}+\mathbf{G}} |\mathrm{e}^{-\mathrm{i}(\mathbf{G}+\mathbf{q})\mathbf{r}}| \phi_{\nu} \rangle\\
  &M_{\nu' \mathbf{G}' \mathbf{k}}(\mathbf{G},\mathbf{q})=\langle \phi_{\nu'} |\mathrm{e}^{-\mathrm{i}(\mathbf{G}+\mathbf{q})\mathbf{r}}| \phi_{\mathbf{k}+\mathbf{q}+\mathbf{G}'} \rangle\\
  &M_{\nu' \nu }(\mathbf{G},\mathbf{q})=\langle \phi_{\nu'} |\mathrm{e}^{-\mathrm{i}(\mathbf{G}+\mathbf{q})\mathbf{r}}| \phi_{\nu} \rangle.
 \end{aligned}
\end{equation}
Equivalently to the case of the momentum-matrix elements, the plane-wave matrix elements are decomposed into an interstitial and a MT part. The integration in the interstitial part is performed analytically. In the MT spheres, we employ the Rayleigh expansion of plane waves in products of spherical harmonics and spherical Bessel functions of first kind $j_l(r)$, \textit{i.e.} $\mathrm{e}^{-\mathrm{i}(\mathbf{G}+\mathbf{q})\mathbf{r}}=4 \pi \sum_{lm}(-\mathrm{i})^l j_l((G+q)r)Y_{lm}(\hat{r})Y_{lm}(\hat{G+q})$.
As in the case of the mometum matrix elements, the spherical integral is performed analytically, while a numerical integration is performed on the radial grid. For plane-wave matrix elements between core and conduction states, a decomposition in the basis is not suitable, and the matrix elements are expressed as
\begin{equation}\label{eq:core-m}
 M_{\kappa M, i \mathbf{k}}(\mathbf{G},\mathbf{q})=4\pi \sum_{lm}\sum_{l'm'}Y_{l'm'}(\hat{G+q})\int d\Omega \; \Omega^{*}_{\kappa,M}(\hat{r})Y_{l'm'}(\hat{r})Y_{lm}(\hat{r})\int_{R_{MT}}dr \; u^{*}_{\kappa,\alpha}(r) j_{l'}((G+q)r)u_{l}^{i \mathbf{k}}(r).
\end{equation}
More details on the calculation of momentum and plane-wave matrix elements in the LAPW+LO basis can be found in Refs.~\cite{vorwerkPRB,Sagmeister}.

We also define modified plane-wave matrix elements $N_{nm \mathbf{k}}(\mathbf{G},\mathbf{q})$ as
\begin{equation}
 N_{mn \mathbf{k}}(\mathbf{G},\mathbf{q})=\langle m \mathbf{k} |\mathrm{e}^{-\mathrm{i}(\mathbf{G}+\mathbf{q})\mathbf{r}}|\left(n (\mathbf{k}+\mathbf{q})\right)^* \rangle,
\end{equation}
which are evaluated as
\begin{equation}
 \begin{aligned}
    N_{mn \mathbf{k}}(\mathbf{G},\mathbf{q})&=\sum_{\mathbf{G}\mathbf{G}'} C_{m (\mathbf{k}+\mathbf{G})}C^*_{n (\mathbf{k}+\mathbf{q}+\mathbf{G})} M_{\mathbf{G}\mathbf{G}'\mathbf{k}}(\mathbf{G},\mathbf{q})+\sum_{\mathbf{G}\nu'} C_{m (\mathbf{k}+\mathbf{G})}C^*_{n \nu' \mathbf{k}} M_{\mathbf{G}\nu' \mathbf{k}}(\mathbf{G},\mathbf{q})\\
    &+\sum_{\nu \mathbf{G}'} C_{m \nu' \mathbf{k}}C^*_{n (\mathbf{k}+\mathbf{q}+\mathbf{G})}M_{\nu \mathbf{G}' \mathbf{k}}(\mathbf{G},\mathbf{q})+\sum_{\nu \nu'} C_{m \nu' \mathbf{k}}C^*_{n \nu \mathbf{k}+\mathbf{q}}M_{\nu \nu' \mathbf{k}}(\mathbf{G},\mathbf{q}).
  \end{aligned}
\end{equation}
For additional details regarding the calculation of the plane-wave matrix elements in the (L)APW+lo basis of \texttt{exciting}, we refer the readers to Refs.~\cite{pusc-ambr02prb,Sagmeister2009}. For additional information on matrix elements between states in the (L)APW+lo basis and core states, we refer to Ref.~\cite{vorwerkPRB}.

\subsection{Matrix elements of the BSE Hamiltonian}
The matrix elements of the exchange interaction in Eqs.~\ref{eqn:diagonalblock} and \ref{eqn:couplingblock} are given by 
\begin{equation}\label{eqn:vij}
  V^{rr}_{ij}(\mathbf{q})=\int \Upsilon^*_{i,\mathbf{q}}(\mathbf{r},\mathbf{r})v(\mathbf{r},\mathbf{r}')\Upsilon_{j,\mathbf{q}}(\mathbf{r}',\mathbf{r}')d^3r d^3r'.
\end{equation}
We introduce the Fourier transform of the bare Coulomb potential 
\begin{equation}
  v(\mathbf{r},\mathbf{r}') = 
  \sum_{\mathbf{G}} \sum_{\mathbf{p}}
  \underbrace{%
  \frac{1}{V_\mathrm{c}}
  \frac{4 \pi}{|\mathbf{G}+\mathbf{p}|}
  }_{=v_\mathbf\mathbf{G}(\mathbf{p})}
  \mathrm{e}^{\mathrm{i} \left( \mathbf{G} + \mathbf{p} \right) \left( \mathbf{r} - \mathbf{r}' \right)},
  \label{eqn:vft}
\end{equation}
where $V_c$ denotes the crystal volume. 
The matrix elements $V^{rr}_{\alpha \alpha'}$ of Eqs.~\ref{eqn:diagonalblock} and \ref{eqn:couplingblock} are computed in reciprocal space as
\begin{equation}
  V^\mathrm{rr}_{\alpha \alpha'}(\mathbf{q}) 
  = \sum_{\mathbf{G}} 
  v_\mathbf{G}(\mathbf{q})
  M^*_{uo\mathbf{k}_-}(\mathbf{G},\mathbf{q})
  M_{u'o'\mathbf{k}_-'}(\mathbf{G},\mathbf{q}).
  \label{eqn:vrrofM}
\end{equation} 
The matrix elements of the screened Coulomb interaction are given by
\begin{equation}
  W_{ij}(\mathbf{q}) =
  \iint \Upsilon^*_{i,\mathbf{q}}(\mathbf{r},\mathbf{r}') 
  w(\mathbf{r},\mathbf{r}')
  \Upsilon_{j,\mathbf{q}}(\mathbf{r},\mathbf{r}') d^3r d^3r'.
  \label{eqn:wij}
\end{equation}
The statically screened Coulomb potential is given by
\begin{equation}
  w(\mathbf{r},\mathbf{r}') = \int v(\mathbf\mathbf{r},\mathbf{r}'') \varepsilon^{-1}(\mathbf{r}'',\mathbf{r}', \omega=0)
  d^3r''.
  \label{eqn:statw}
\end{equation}
Again, we make use of its Fourier representation  
\begin{equation}
  w(\mathbf{r},\mathbf{r}') = 
  \sum_{\mathbf{G} \mathbf{G}'} \sum_{\mathbf{p}}
  \mathrm{e}^{\mathrm{i} \left( \mathbf{G} + \mathbf{p} \right) \mathbf{r}}
  \,
  w_{\mathbf{G}\mathbf\mathbf{G}'}(\mathbf{p}, \omega=0)
  \,
  \mathrm{e}^{-\mathrm{i} \left( \mathbf{G}' + \mathbf{p} \right) \mathbf{r}'},
  \label{eqn:wft}
\end{equation}
where the Fourier components are given by
\begin{equation}
  w_{\mathbf{G} \mathbf{G}'}(\mathbf{p}) = 
  v_{\mathbf{G}}(\mathbf{p}) \varepsilon^{-1}_{\mathbf{G}
  \mathbf{G}'}( \mathbf{p}, \omega=0).
  \label{eqn:wggq}
\end{equation}
Here, the dielectric function is computed in the random-phase approximation (RPA) $\varepsilon_{\mathbf{G}\mathbf{G}'}(\mathbf{q},\omega) \approx \varepsilon^{RPA}_{\mathbf{G}\mathbf{G}'}(\mathbf{q},\omega)$ \cite{Ehrenreich59} as
\begin{equation}
	\varepsilon^{RPA}_{\mathbf{G}\mathbf{G}'}(\mathbf{q}, \omega)=\delta_{\mathbf{G}\mathbf{G}'}-\frac{1}{V_c}v_{\mathbf{G}'}(\mathbf{q})\sum_{ij \mathbf{k}} \frac{f(\epsilon_{j \mathbf{k}+\mathbf{q}})-f(\epsilon_{i \mathbf{k}})}{\epsilon_{j \mathbf{k}+\mathbf{q}}-\epsilon_{i \mathbf{k}}-\omega}\left[ M_{ij}^{\mathbf{G}}(\mathbf{k},\mathbf{q}) \right]^* M_{ij}^{\mathbf{G}'}(\mathbf{k},\mathbf{q}),
\end{equation}
where $f(\epsilon_{i \mathbf{k}})$ are the occupation factors of the single-particle state with energy $\epsilon_{i \mathbf{k}}$.
In terms of the plane-wave matrix elements, the resonant-resonant block of Eq.~\eqref{eqn:wij} can be rewritten as 
\begin{equation}
  W^\mathrm{rr}_{\alpha \alpha'}(\mathbf{q})=
  \sum_{\mathbf{G}\mathbf{G}'} w_{\mathbf{G} \mathbf{G}'}( \mathbf{k}-\mathbf{k}')
  M^*_{o' o \mathbf{k}'_+}(\mathbf{G}, \mathbf{k}-\mathbf{k}')
  M_{u' u \mathbf{k}'_-}(\mathbf{G}', \mathbf{k}-\mathbf{k}').
  \label{eqn:wrrm}
\end{equation}
The  elements of the resonant-anti-resonant block can be computed as
\begin{equation}
  W^{\mathrm{ra}}_{\alpha \alpha'}(\mathbf{q})
  = \frac{1}{V} \sum_{\mathbf{G}\mathbf\mathbf{G}'} w_{\mathbf{G}\mathbf{G}'}(-\mathbf{k'}-\mathbf{k})
  N^*_{u o' \mathbf{k}_-}(\mathbf{G}, -\mathbf{k'}-\mathbf{k})
  N_{o u' \mathbf{k}_+}(\mathbf{G}', -\mathbf{k'}-\mathbf{k}),
  \label{eqn:wrabar}
\end{equation}

\subsection{BSE as an eigenvalue problem}
The resolvent $L(\mathbf{q}, \omega)=-\left[ \mathrm{H}(\mathbf{q})-\omega \mathrm{\Delta}\right]^{-1}$ of Eq.~\eqref{eqn:chiofh} can be found using the solutions of the generalized eigenvalue problem (the index $\mathbf{q}$ is dropped for simplicity)
\begin{equation}
  \mathrm{H} \mqty(\mathbf{X}_\lambda \\ \mathbf{Y}_\lambda) = 
  E_\lambda \mathrm{\Delta} \mqty(\mathbf{X}_\lambda \\ \mathbf{Y}_\lambda),
  \label{eqn:gevpbse}
\end{equation}
where, according to Ref.~\cite{Furche_2001},
\begin{equation}
  \left[ \mathrm{H}- \omega \mathrm{\Delta}\right]^{-1} =\sum_\lambda\frac{1}{E_\lambda-\omega} 
  \mqty(\mathbf{X}_\lambda \\ \mathbf{Y}_\lambda) \mqty(\mathbf{X}_\lambda \\ \mathbf{Y}_\lambda)^\dag + \frac{1}{E_\lambda+\omega}
  \mqty(\mathbf{Y}_\lambda \\ \mathbf{X}_\lambda) \mqty(\mathbf{Y}_\lambda \\ \mathbf{X}_\lambda)^\dag.
  \label{eqn:resolventwithev}
\end{equation}
For the solution of the full BSE, a direct diagonalization scheme is adopted in \texttt{exciting}.
This scheme~\cite{Sander2015, Furche_2001} maps the generalized eigenvalue problem of Eq.~\eqref{eqn:gevpbse} onto an auxiliary  eigenvalue problem of half its size. 
The auxiliary Hamiltonian is constructed as
\begin{equation}
  \mathrm{S} = 
  \left( \mathrm{A} - \mathrm{B} \right)^{\frac{1}{2}}
  \left( \mathrm{A} + \mathrm{B} \right)
  \left( \mathrm{A} - \mathrm{B} \right)^{\frac{1}{2}},
  \label{eqn:auxham}
\end{equation}
and the solutions of
\begin{equation}
  \mathrm{S} \mathbf{Z}_\lambda = E^2_\lambda \mathbf{Z}_\lambda
  \label{eqn:auxevp}
\end{equation}
are used to reconstruct eivenvalues and eigenvectors of Eq.~\eqref{eqn:gevpbse}. As long as $\mathrm{A}-\mathrm{B}$ and $\mathrm{A}+\mathrm{B}$ are positive definite, the solutions of Eq.~\ref{eqn:gevpbse} are given by
 \begin{equation}
     \mathbf{X}_{\lambda}+\mathbf{Y}_{\lambda} =\left( \mathrm{A}-\mathrm{B} \right)^{\frac{1}{2}}\frac{1}{\sqrt{E_{\lambda}}}\mathbf{Z_{\lambda}}
 \end{equation}
and
 \begin{equation}
    \mathbf{X}_{\lambda}-\mathbf{Y}_{\lambda} =\left( \mathrm{A}-\mathrm{B} \right)^{\frac{1}{2}}\sqrt{E_{\lambda}}\mathbf{Z_{\lambda}}.
 \end{equation}
In the TDA, the coupling blocks between the resonant and anti-resonant subspace are neglected, namely $\mathrm{B} = 0$, and the eigenvalue problem 
\begin{equation}
  \mathrm{H}^\mathrm{TDA} \mathbf{X}_\lambda = \mathrm{A} \mathbf{X}_\lambda = E_\lambda \mathbf{X}_\lambda
  \label{eqn:tdaevp}
\end{equation}
is solved.

\subsection{Dielectric properties from the solutions of the BSE}
The polarizability $P_{\mathbf{G}\mathbf{G}'}(\mathbf{q},\omega)$ is obtained as
\begin{equation}
  P_{\mathbf{G}\mathbf{G}'}(\mathbf{q},\omega) = 
  \frac{1}{V_\mathrm{c}}
  \sum_{i j}
  M_{\alpha}(\mathbf{G}, \mathbf{q})
  L_{i j}(\omega)
  M^*_{\beta}(\mathbf{G}', \mathbf{q})
  \label{eqn:pbargg}
\end{equation}
from the eigenstates of the BSE Hamiltonian through the matrix elements $L_{\alpha \beta}(\mathbf{q},\omega)$ of Eq.~\eqref{eqn:chiofh} and the plane wave matrix elements.
The plane-wave matrix elements are given as $M_\alpha(\mathbf{G},\mathbf{q})=M_{uo\mathbf{k}_-}(\mathbf{G},\mathbf{q})$.
By using Eqs.~\eqref{eqn:chiofh} and \eqref{eqn:resolventwithev}, Eq.~\eqref{eqn:pbargg} can be written in terms of the BSE eigenvalues and eigenvectors as follows
\begin{equation}
  P_{\mathbf{G}\mathbf{G}'}(\mathbf{q},\omega) 
  =\sum_\lambda
  \left(
  \frac{1}{\omega - E_\lambda +\mathrm{i} \delta} + \frac{1}{-\omega - E_\lambda - \mathrm{i} \delta}
  \right)
  t^*_\lambda(\mathbf{G},\mathbf{q})
  t_\lambda(\mathbf{G}',\mathbf{q}),
  \label{eqn:chir}
\end{equation}
where we have introduced the transition coefficients $t_\lambda$ given by:
\begin{equation}
  t_\lambda(\mathbf{G},\mathbf{q}) = 
  V^{-\frac{1}{2}}_\mathrm{c} \sum_\alpha \left( \mathbf{X}_\lambda + \mathbf{Y}_\lambda \right)^\dag_\alpha 
  \mathbf{M}^{*}_\alpha(\mathbf{G},\mathbf{q}).
  \label{eqn:osci}
\end{equation}
These terms represent a sum of weighted plane-wave transition matrix elements, where the weights are computed from the corresponding BSE eigenvectors.
The signs of the imaginary broadening factors $\mathrm{i} \delta$ are chosen such that the resulting response function is retarded (see also Ref.~\cite{Strinati1988}).
We then obtain the macroscopic dielectric function $\varepsilon_M(\mathbf{G}+\mathbf{q},\omega)$ as
\begin{equation}
 \varepsilon_M(\mathbf{G}+\mathbf{q}, \omega)=\frac{1}{1+v_{\mathbf{G}}(\mathbf{q})P_{\mathbf{G}\mathbf{G}}(\mathbf{q},\omega)}
\end{equation}

In the optical limit, \textit{i.e.} $\mathbf{G}=0$, $\mathbf{G}'=0$, and $\mathbf{q}_\mathrm{mt} \rightarrow 0$, the divergence of the bare Coulomb potential has to be addressed. To do so, the plane-wave matrix elements $M$ in $\mathbf{q}$ are expanded in $\mathbf{q}$ to first order, and an expression in terms of the dipole matrix elements $D$ is obtained
\begin{equation}
  \lim_{\mathbf{q}\rightarrow 0} M_{\alpha}(\mathbf{G}=0, \mathbf{q})= \mathrm{i} \sum_{j}q_{j} D_{\alpha,j},
  \label{eqn:mofd}
\end{equation}
where $D_{\alpha,j}$ is defined in terms of the momentum operator
\begin{equation}
  D_{\alpha,j}
  = \mathrm{i} \frac{\matrixel{u \mathbf{k}}{\hat{p}_j}{o  \mathbf{k}}}
  {\epsilon_{u  \mathbf{k}} - \epsilon_{o \mathbf{k}}}.
  \label{eqn:dipolfromp}
\end{equation}
Inserting Eq.~\eqref{eqn:mofd} into Eq.~\eqref{eqn:osci} yields 
\begin{equation}
  t_\lambda(0,\mathbf{q}) = -\mathrm{i} \frac{\hat{\mathbf{q}}}{\abs{\mathbf{q}}} \left( \mathbf{X}_\lambda + \mathbf{Y}_\lambda \right)^\dag \mathrm{D}^{*},
  \label{eqn:osciwithq}
\end{equation}
where $\hat{\mathbf{q}}$ is the unit vector along the direction of $\mathbf{q}$.
In this way the factor $1/\abs{\mathbf{q}}^2$ of the Coulomb potential is canceled out. We obtain the macroscopic dielectric function in the optical limit as
\begin{equation}
 \varepsilon_M(\mathbf\mathbf{q}\rightarrow 0, \omega)=\frac{1}{\lim_{\mathbf{q}\rightarrow 0} \varepsilon^{-1}_{00}(\mathbf{q},\omega)}.
\end{equation}
Alternatively, an \textit{effective} polarizability $\bar{P}_{\mathbf{G}\mathbf{G}'}$ is employed to obtain the macroscopic dielectric function directly:
\begin{equation}
 \varepsilon_{\mathrm{M}}(\mathbf{G}+\mathbf{q},\omega) =1 - \frac{4\pi}{|\mathbf{G}+\mathbf{q}|^2} \bar{P}_{\mathbf{G}\mathbf{G}}(\mathbf{q},\omega).
\end{equation}
The effective polarizability $\bar{P}$ is connected to the polarizability $P$ via $\bar{P}=P+P\bar{v}\bar{P}$,
where $\bar{v}$ is given for a momentum transfer $\mathbf{Q}=\mathbf{G}+\mathbf{q}$ as
\begin{equation}
  \bar{v}_{\mathbf{G}'}(\mathbf{q}')=\begin{cases} 0 & \mathbf{G}'=\mathbf{G} \;\&\; \mathbf{q}'=\mathbf{q}\\ v_{\mathbf{G}'}(\mathbf{q}') & \mathrm{else}\end{cases}
  \label{eqn:vbar_def}
\end{equation}
The effective polarizability is calculated by simply replacing the bare Coulomb potential in Eqs.~\ref{eqn:vij} and \ref{eqn:vrrofM} with $\bar{v}$. More details can be found in Refs.~\cite{Onida2002,Sander2015,aurich-thesis2017}. In the optical limit, this yields the dielectric function as
\begin{equation}
  \begin{split}
    \varepsilon_{\mathrm{M}}(\omega) &=1 - \lim_{q \rightarrow 0} \frac{4\pi}{q^2} \bar{P}_{0,0}(\mathbf{q},\omega) 
    \\&=1 -4 \pi \sum_{i,j}\hat{q}_i \hat{q}_j \sum_\lambda \left(\frac{t^*_{\lambda,i}t_{\lambda,j}}{\omega-E_\lambda+\mathrm{i} \delta} + \frac{t^*_{\lambda,i}t_{\lambda,j}}{-\omega-E_\lambda-\mathrm{i} \delta}\right),
 \end{split}
  \label{eqn:epsmtensor}
\end{equation}
where 
\begin{equation}
  t_{\lambda,i} 
  = 
  -\mathrm{i} \sum_{\alpha}
  \left( \mathbf{X}_\lambda + \mathbf{Y}_\lambda \right)^\dag_\alpha 
  \mathrm{\tilde{D}}^{*}_{\alpha,i}
  \label{eqn:tlambdai}
\end{equation}
define the transition coefficients for each Cartesian direction.
From this expression we can recover the form of the (macroscopic) transversal dielectric matrix of Eq.~\eqref{eqn:epsmofpbar},
\begin{equation}
  \varepsilon^{ij}_{\mathrm{M}}(\omega) = 
  \delta_{ij} - 4\pi 
  \sum_\lambda
  \left( \frac{t^*_{\lambda,i}t_{\lambda,j}}{\omega-E_\lambda+\mathrm{i} \delta} + \frac{t^*_{\lambda,i}t_{\lambda,j}}{-\omega-E_\lambda-\mathrm{i}\mathrm{i} \delta}
  \right).
  \label{eqn:epsmij}
\end{equation}
In scattering spectroscopies, the double-differential cross-section $\pdv{\sigma}{\Omega}(\mathbf{Q})$ is measured, which determines the number of particles (electrons or x-ray photons, respectively) scattered into the solid angle element $\Omega$ and which loose energy $\hbar \omega$ and momentum $\mathbf{Q}$. The double-differential cross-section of IXS and EELS measurements are related to the macroscopic dielectric function as \cite{Kubo1966a,hambach}

%
\begin{equation}
  \pdv{\sigma}{\Omega}{\omega}(\mathbf{Q}=\mathbf{G}+\mathbf{q}) \propto
  -\frac{1}{\pi} 
  v^{-1}_{\mathbf{G}}(\mathbf{q}) 
  \Im\frac{1}{\varepsilon_{M}
  (\mathbf{q},\omega)},
\end{equation}
where $v^{-1}_{\mathbf{G}}(\mathbf{q})$ is the Coulomb potential, the momentum loss $\mathbf{Q}$ is decomposed into a reciprocal lattice vector $\mathbf{G}$ and a vector $\mathbf{q}$ of the first Brillouin zone. Scattering spectra are often compared to the \textit{dynamical structure factor} $S(\mathbf{Q},\omega)$ which is defined as
\begin{equation}
  S(\mathbf{Q}=\mathbf{G}+\mathbf{q},\omega) =
  -\frac{1}{\pi} 
  v^{-1}_{\mathbf{G}}(\mathbf{q}) 
  \Im \frac{1}{\varepsilon_{M}
  (\mathbf{q},\omega)},
\end{equation}
and also to the \textit{electron energy-loss function} $\mathcal{L}$, defined as
\begin{equation}
  \mathcal{L}(\mathbf{Q}=\mathbf{G}+\mathbf{q},\omega) = -\Im\frac{1}{\varepsilon_{M}(\mathbf{q},\omega)}.
  \label{eqn:Lofeps} 
\end{equation}
%
\subsection{BSE calculations for core spectroscopy}
The procedure described above naturally applies also in the case of core spectroscopy, when the BSE is solved for transitions between core and conduction states.
In this case, the initial states are naturally selected among the core levels of Eq.~\ref{eq:core} that are obtained in the (L)APW+lo formalism of \texttt{exciting} as solutions of the radial Dirac equation. While the implementation of the momentum and plane-wave matrix elements of Eqs.~\ref{eq:core-p} and~\eqref{eq:core-m} are modified to include core states, the subsuquent calculation of the dielectric properties is performed analogously to the one in the optical region.
More details about the determination of core states and the calculation of core-conduction matrix elements are given in Ref.~\cite{vorwerkPRB}.
\begin{figure}[h]
\center
  \includegraphics[width=.4\textwidth]{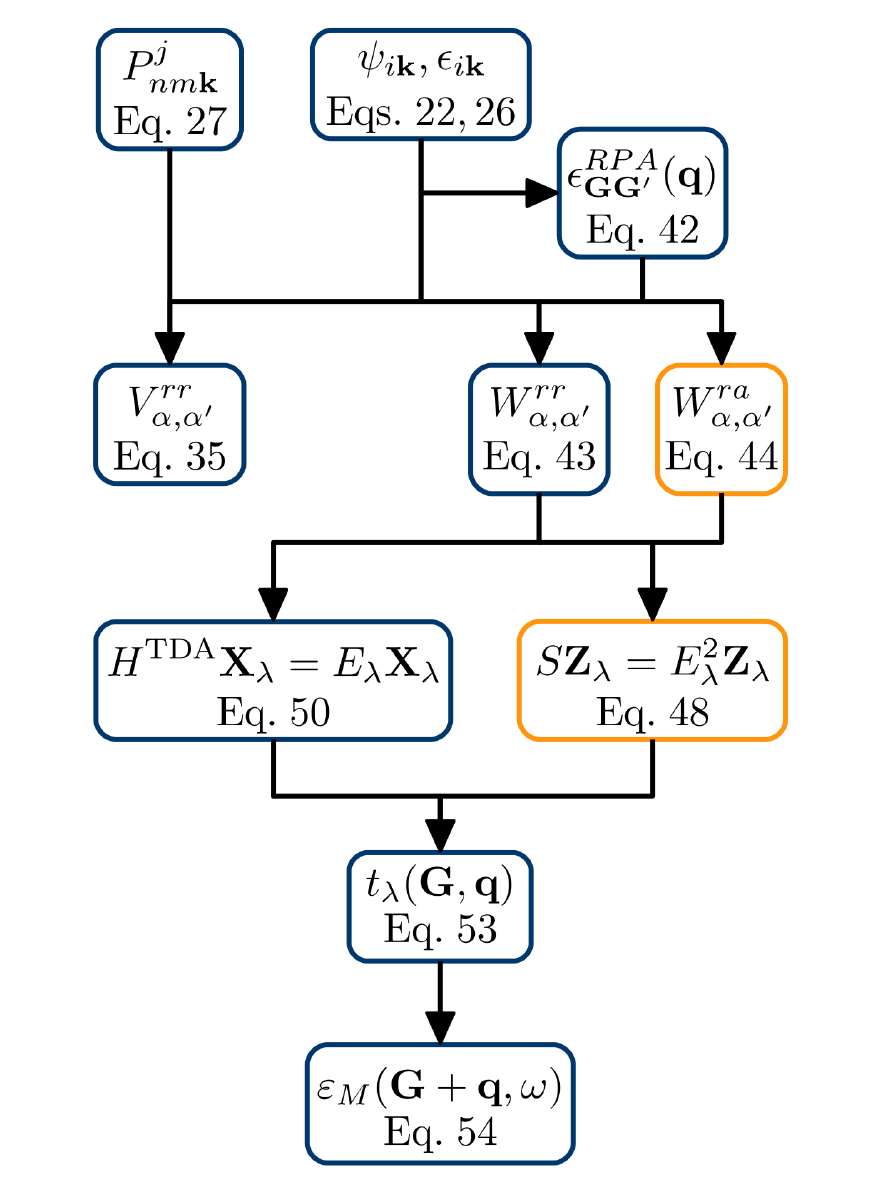}
  \caption{Schematic workflow of BSE calculations in \texttt{exciting}. Quantities that are calculated if the Tamm-Dancoff approximation is lifted, are shown in yellow.}
  \label{fig:workflow}
\end{figure}
%
\section{Applications}
In this section, we present selected applications of the methodology illustrated above and implemented in the \texttt{exciting} code.
In Sec.~\ref{sec:LiF} we show the results of $\mathbf{q}$-dependent BSE to describe the dynamical structure factor, the exciton band structure and the core spectra of LiF, a crystalline insulator.
To demonstrate the importance to go beyond the TDA, we analyze the loss function of bulk silicon (Sec.~\ref{sec:Si}) and the optical absorption of a biphenyl crystal and of monolayers of trans- and cis-azobenzene molecules (Sec.~\ref{sec:organics}).

\subsection{Dynamical structure factor}
\label{sec:LiF}

\begin{figure}
\center
  \includegraphics{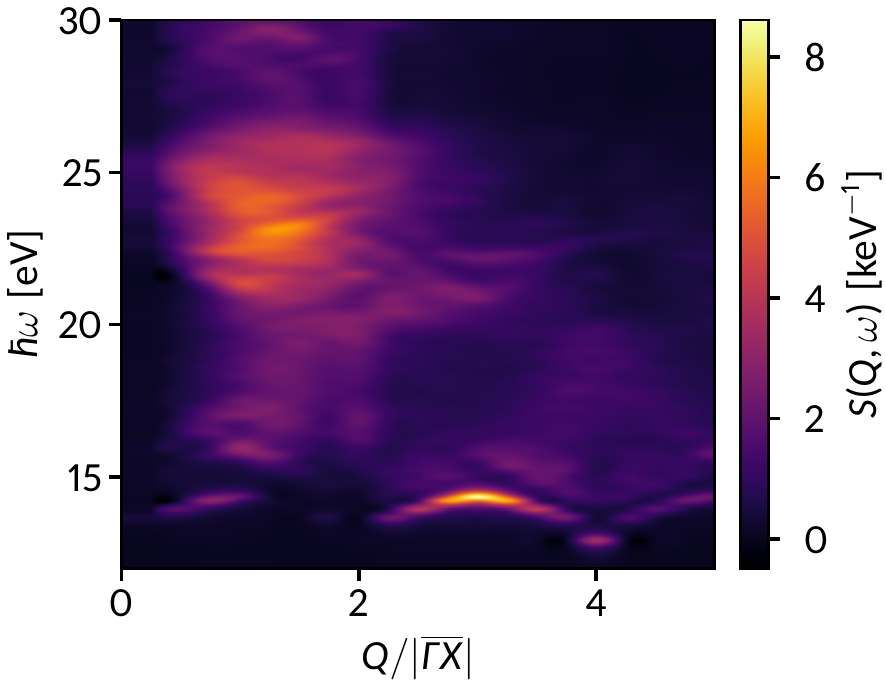}
  \caption{Dynamical structure factor of LiF as a function of the energy loss (vertical axis) and the momentum loss (horizontal axis). The momentum loss is chosen along the $\Gamma$--$X$ direction until the border of the 5th Brillouin zone.}
  \label{fig:LiFGXcol}
\end{figure}
%

In order to reproduce and interpret EELS and IXS experiments, the BSE formalism needs to be considered at finite momentum transfer. To demonstrate the capability of our implementation, we determine the dynamical structure factor of LiF in the optical loss region. Calculations are performed along the $\Gamma$--$X$ path of the BZ until the border of the 5th Brillouin zone (see Fig.~\ref{fig:LiFGXcol}), corresponding to transferred momentum $\mathbf{q_x}$, with $\mathbf{q_y} = \mathbf{q_z}=$ 0.
For this calculation, a shifted 8 $\times$ 8 $\times$ 8  $\mathbf{k}$-mesh is employed.
The transition space is formed by 4 occupied and 12 unoccupied bands, and 50 empty states are included in the calculation of the screened Coulomb interaction.
Local-field effects are taken into account up to a cutoff of $|\mathbf{G} + \mathbf{q}|_{max} = 5\; a_0^{-1}$.
\begin{figure}
\center
  \includegraphics[width=.5\textwidth]{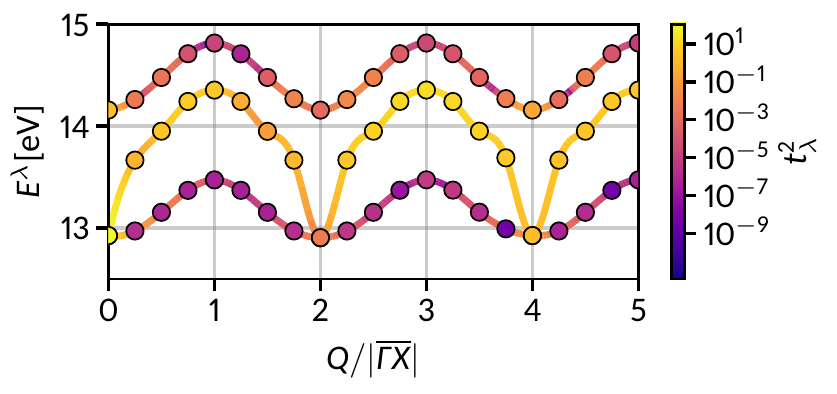}
  \caption{Exciton band structure of LiF along the $\Gamma$--$X$ path for the three lowest energy excitons of the spectrum in Fig.~\ref{fig:LiFGXcol}. Oscillator strengths are indicated according to the color scheme shown by the color bar.}
  \label{fig:LiFGXexdisp}
\end{figure}

The dynamical structure factor of LiF is shown in Fig.~\ref{fig:LiFGXcol} over a broad energy window between 12.5 and 30 eV, where the spectrum of this insulating material exhibits its most relevant features~\cite{roes-walk67josa}. 
The low-energy region is characterized by a tightly bound exciton at about 14 eV, which gives rise to a narrow peak with a significant dispersion starting from 2$\overline{\Gamma X}$ up to about 5$\overline{\Gamma X}$.
The intensity of this excitation is modulated, with a pronounced maximum at 3$\overline{\Gamma X}$.
At higher energies and at low $\mathbf{q}$, $S(Q,\omega)$ becomes large between 20 and 25 eV, with a maximum at about 23 eV.
This broad feature, which exhibits dispersion between 1$\overline{\Gamma }$ and 2$\overline{\Gamma X}$, is related to the plasmon peak, as discussed in Refs.~\cite{gatt-sott13prb,abba+08pnas}.
It shifts to lower energies upon increasing $\mathbf{q}$ and eventually vanishes at around 3.5 $\overline{\Gamma X}$.
The ultimate disappearance of plasmon and exciton peaks at high $\mathbf{q}$-values is ascribed to the decay of the plane-wave matrix elements~\cite{gatt-sott13prb}. 
Our result is in excellent agreement with the findings of a previous study at the same level of theory~\cite{gatt-sott13prb} and with recent experimental results on the dynamical structure factor of LiF~\cite{celi+00prl,abba+08pnas}.

\subsection{Exciton band structure}
Excitonic bandstructures, \textit{i.e.} the dispersion of exciton binding energies with momentum loss, are obtained from the full diagonalization of the BSE Hamiltonian at finite momentum transfer. This bandstructure ammends the $\mathbf{q}$-dependent dielectric properties obtained from the calculations.
Exemplary, in Fig.~\ref{fig:LiFGXcol}, we report the dispersion associated to the first three (bound) excitons in the optical spectrum of LiF along the $\Gamma$–X direction.
It can be immediately noticed that exciton energies are periodic with respect to the reciprocal lattice vector 2$\overline{\Gamma X}$.
The lowest-energy exciton is two-fold degenerate along the entire momentum-transfer path considered in these calculations. 
The third exciton is degenerate with the first two at $\Gamma$ and at equivalent points, but upon finite transferred momentum it exhibits a different dispersion reaching higher energies.
At $\Gamma$ the fourth exciton is energetically higher by more than 1 eV compared to the first three.
Its dispersion follows the modulation of the lowest-energy branch with maxima (minima) at odd (even) multiples of $\mathbf{q}$ along the $\Gamma$–X direction.
Absolute exciton energies increase with the distance from the $\Gamma$-point (or any equivalent point), due to the direct nature of the band gap in LiF at $\Gamma$.
Excitons that are not composed of vertical transitions have necessarily higher energies than those formed by vertical ones. 
The oscillator strength of these excitations varies with and overall decays with respect to $\mathbf{q}$. 

\subsection{Core spectroscopy at finite momentum transfer}
\begin{figure}
\center
\includegraphics[width=.45\textwidth]{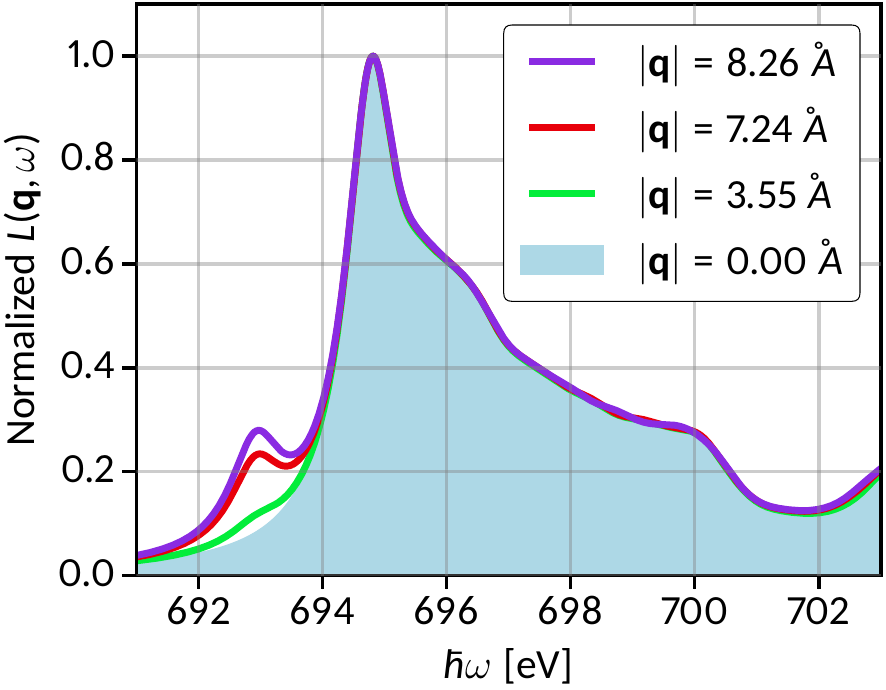}%
\caption{Normalized loss function for the F K edge of LiF with increasing values of momentum transfer. All spectra are normalized at the main peak. Spectra are broadened by a Lorentzian function with full-width at half maximum of 0.5 eV.}
\label{fig:K-Edge}
\end{figure}

Our implementation allows us to treat optical and core excitations on the same footing. As an example, we discuss the loss function of LiF at the flourine K-edge (F 1$s$ electrons are excited) for different values of the momentum $\mathbf{q}$ along the $\Gamma-X$ path. The results of these calculations are shown in Fig.~~\ref{fig:K-Edge}.
Calculations are performed on a shifted $\mathbf{k}$-grid with 13 $\times$ 13 $\times$ 13 points. 
The transition space consists of the 2 occupied F $1s$ states and 20 unoccupied states in the conduction region. 
100 empty states are included in the RPA calculation of the screened Coulomb interaction. The calculated spectrum is shifted by 38.2 eV, such that the main peak of the spectra is aligned with the same feature in the experimental spectrum of Ref.~\cite{hamal+02PRB}.

All spectra shown in Fig.~\ref{fig:K-Edge} display a main peak at approximately 695 eV and perfectly overlap at higher energies. 
At about 693 eV, an excitonic pre-peak appears for $\mathbf{q}>0$, the oscillator strength of which is increasing with the transferred momentum.
This bound exciton, with a binding energy of $\sim$3 eV, is already present for $\mathbf{q}=0$, but it is not visible being dipole-forbidden. 
At finite $\mathbf{q}$ values, the dipole selection rules do not apply, and the peak gains intensity. 
Our calculations are in good agreement with measurements~\cite{hamal+02PRB} and previous \textit{ab initio} calculations~\cite{vinson+11PRB,joly+17JCTC,schw+17jesrp}.

\subsection{Loss function beyond the Tamm-Dancoff approximation}
\label{sec:Si}
As an example for the influence of the Tamm-Dancoff approximation on the loss function of semiconductors, we consider the loss function of bulk silicon. In this calculation, the transition space is formed by 4 occupied and 12 unoccupied bands on a shifted 8 $\times$ 8 $\times$ 8 $\mathbf{k}$-grid.
100 empty bands are included in the RPA calculation of the screened Coulomb interaction.
A scissors shift of 0.95 eV is applied in order to mimic the quasi-particle correction. 
Local-field effects are included with a cutoff of $|\mathbf{G} + \mathbf{q}|_{max} = 5 \; a_0^{-1}$.
BSE calculations are performed by solving the full Hamiltonian and, for comparison, by applying the TDA.

\begin{figure}
\center
\includegraphics[width=.45\textwidth]{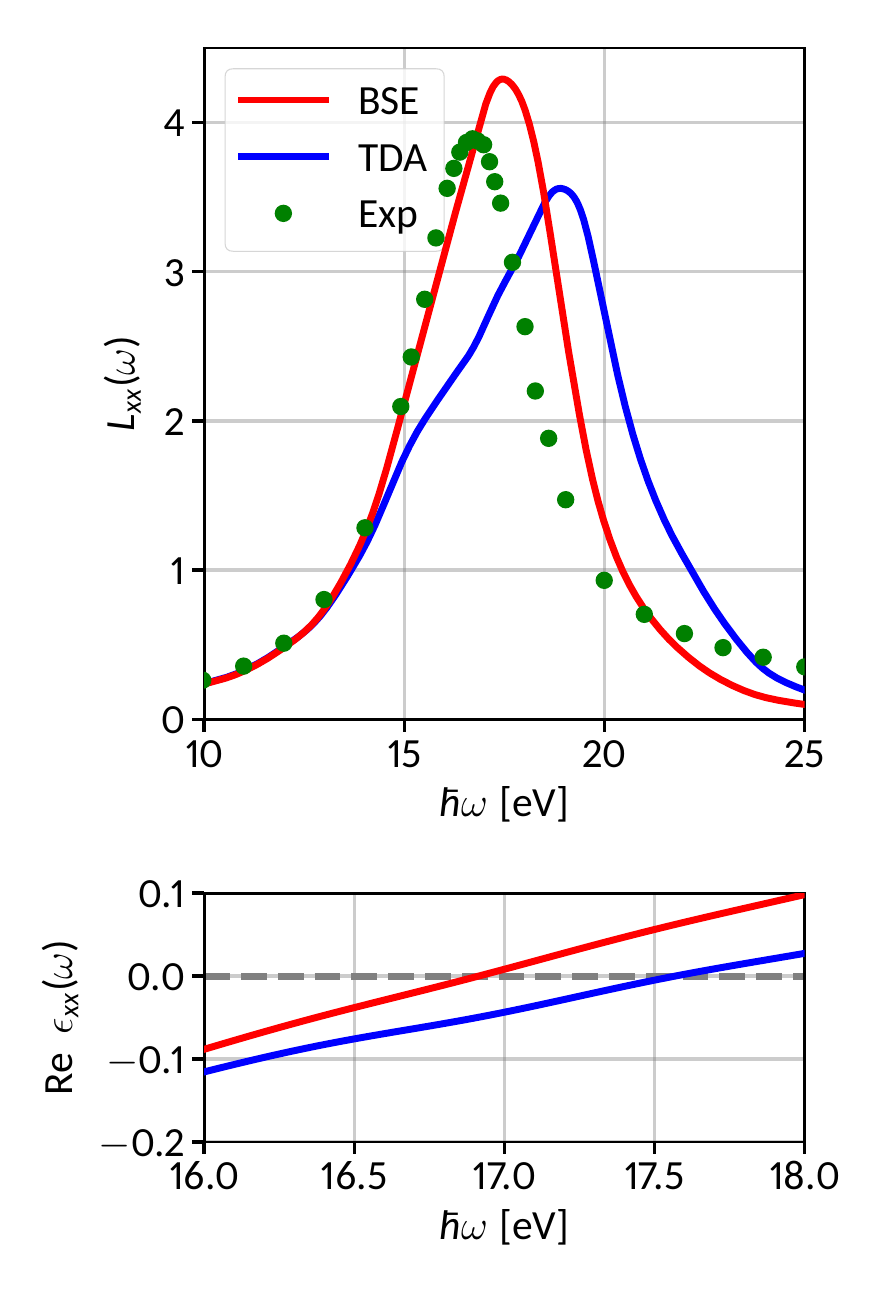}%
\caption{Top: Loss function of bulk silicon computed from the full BSE (red) and in the TDA (blue) and compared to the experimental results from Ref.~\cite{stie78zpb}. All curves are broadened by a Lorentzian function with full-width at half maximum of 0.1 eV. Bottom: Real part of the dielectric function in the vicinity of the energy where it changes sign from negative to positive, corresponding to the maximum of the loss function shown in the top panel. The full BSE result is shown in red, the one obtained within the TDA in blue.}
\label{fig:SiLoss0}
\end{figure}
The calculated loss function of silicon is shown in Fig.~\ref{fig:SiLoss0} (top panel). It exhibits a pronounced peak at approximately 16 eV, which corresponds to a plasmonic resonance.
At the frequency where the loss function has its maximum, \textit{i.e.} the plasmon frequency, the real part of vanishes, as shown in the bottom panel.
Considerable differences emerge between the calculations performed with and without the TDA, as extensively discussed in Ref.~\cite{olev-rein01prl}.
In the TDA, where the coupling between transitions at positive and negative frequencies is neglected, the position of the peak maximum is overestimated by more than 1 eV compared to the experimental results from Ref.~\cite{stie78zpb}.
On the other hand, when the TDA is lifted, resonant and anti-resonant transitions are appropriately coupled and the resulting plasmon peak in the EELS is in very good agreement with the experimental one.
As discussed in Ref.~\cite{olev-rein01prl}, the underlying physical mechanism is the coupling of the plasmon resonance with excitonic effects, which is appropriately reproduced only going beyond the TDA.

\subsection{Optical spectra beyond the Tamm-Dancoff approximation}
\label{sec:organics}
\begin{figure}
\center
\includegraphics{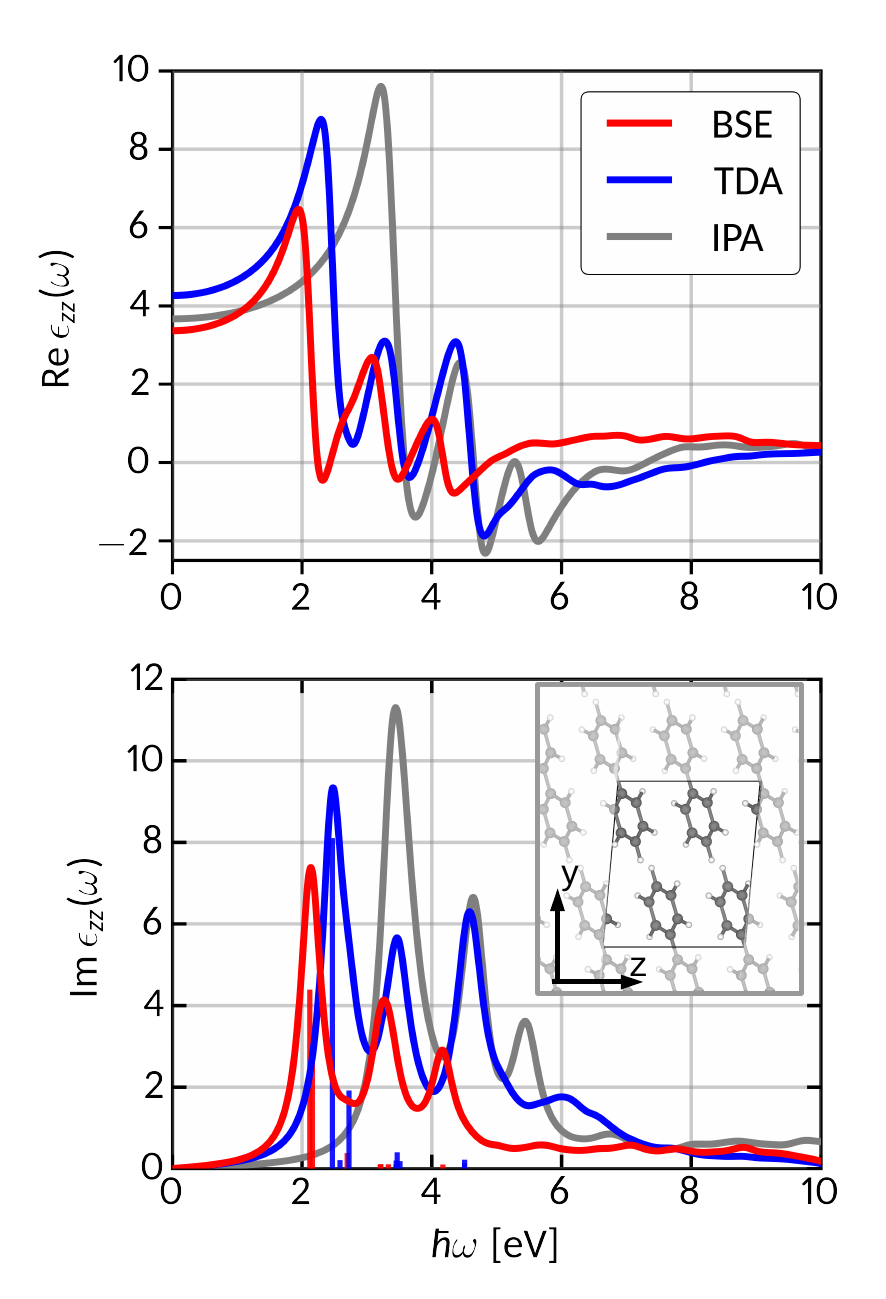}%
\caption{Real (top) and imaginary (bottom) part of the $zz$-component of the macroscopic dielectric tensor of biphenyl crystal, sketched in the inset of the left panel. Results for full BSE (red), BSE in the TDA (blue) and IPA (gray). The vertical bars in the bottom panel indicate the energy and relative oscillator strength of the most dominant excitations. Spectra are broadened by a Lorentzian function with full-width at half maximum of 0.2 eV.}
\label{fig:2Peps}
\end{figure}
The effects of the TDA in optical absorption spectra are expected to be pronounced in organic materials, where the exciton binding energy is a sizable fraction of the bandgap~\cite{pusc+13condmat}. 
In molecular crystals constituted by small molecules like biphenyl this effect is significant.
In Fig.~\ref{fig:2Peps}, we show the real (top panel) and imaginary part (bottom panel) of the macroscopic dielectric function calculated with and without the TDA.
The result obtained in the independent-particle approximation (IPA) is shown for comparison. 
These calculations are performed using a 9 $\times$ 6 $\times$ 5 $\mathbf{k}$-mesh on a transition space including 24 occupied and 13 unoccupied bands.
50 empty states are adopted in the RPA calculation for determining the screened Coulomb interaction and local-field effects are included with a cutoff of $|\mathbf{G} + \mathbf{q}|_{max} = 2.5 \; a_0^{-1}$.

The spectrum exhibits pronounced excitonic effects, which red-shift the absorption onset by more than 1 eV compared to the IPA spectrum.
Going beyond the TDA further decreases the absorption maximum by approximately 0.2 eV.
We also notice a redistribution of the oscillator strength such that the intensity of the lower-energy peaks is lower compared to their TDA counterpart.
While at high energies $\Re \epsilon_M$ converges towards the same value no matter whether the TDA is applied or not, at vanishing frequencies the full BSE yields a lower value of $\Re \epsilon_M$ compared to both IPA and TDA calculations.
This suggests that the missing coupling between excitations and de-excitations in the TDA tends to slightly overestimate the screening, as indicated also by the lower binding energy (higher excitation energy) of the first intense peak.

\begin{figure}
\center
  \includegraphics[width=.45\textwidth]{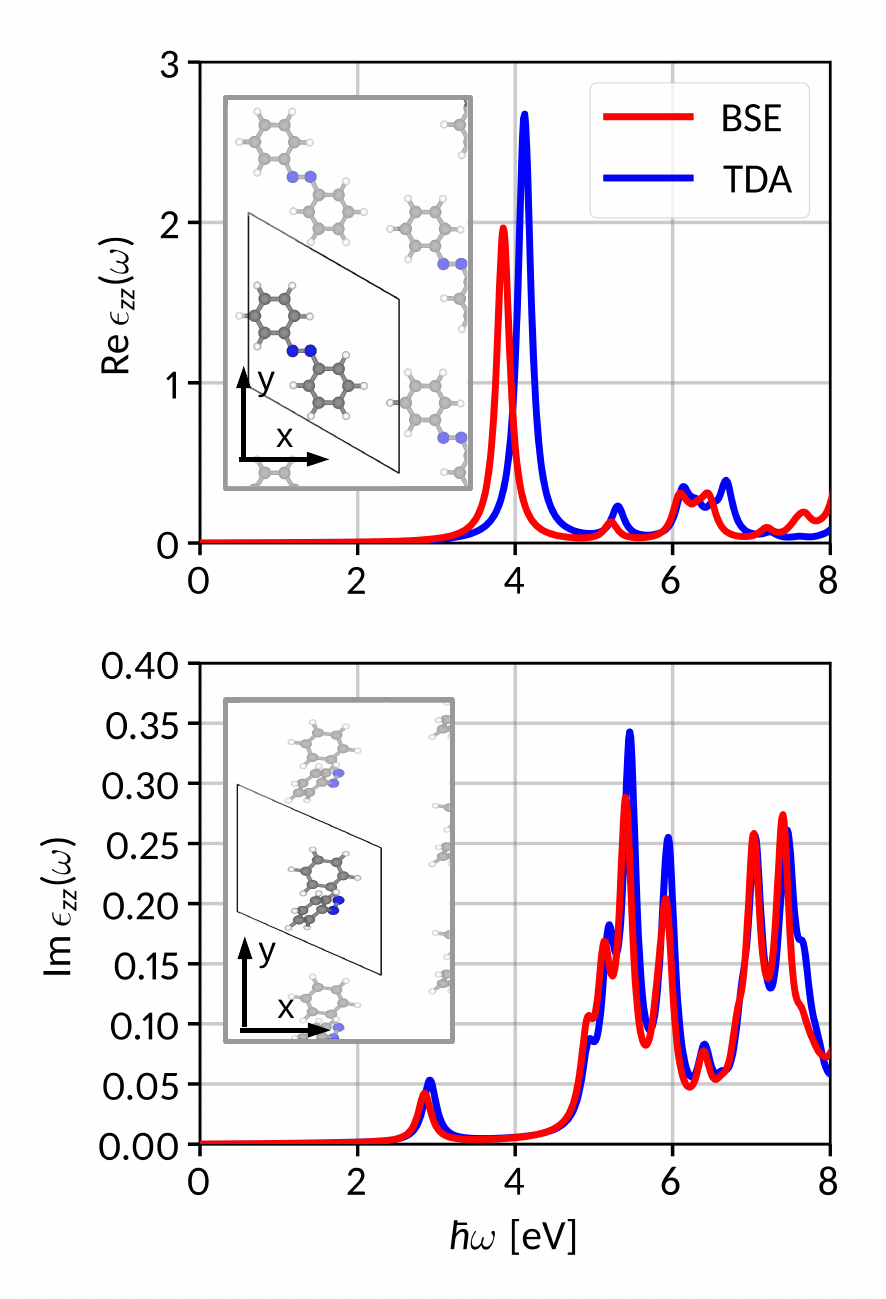}
  \caption{Optical absorption spectrum of a trans-azobenzene monolayer (top) and its counterpart in the cis-phase (bottom) given by the $xx$ component of the imaginary part of the macroscopic dielectric function. Full BSE results are shown in red, the TDA in blue, and the IP ones in gray. The systems and their unit cells are sketched in the insets. Spectra are broadened by a Lorentzian function with full-width at half maximum of 0.2 eV.}
  \label{fig:transazo}
\end{figure}

As another example for the effects of the TDA in organic materials, we consider the optical absorption spectra of monolayers of trans- and cis-azobenzene molecules (see Fig.~\ref{fig:transazo}). 
For these calculations, the BZ is sampled by a 4 $\times$ 4 $\times$ 1 $\mathbf{k}$-mesh.
Transitions between the highest 12 occupied bands and the lowest 12 unoccupied bands are considered.
300 empty states are included in the RPA calculation of the screened Coulomb interaction.
Optical excitations of both trans- and cis-azobenzene monolayers were recently studied from MBPT, by solving the BSE within the TDA~\cite{fu+17pccp}.
Here, we aim to understand the effect of the TDA on the position of the peaks and on the overall spectral shape.

In the spectrum of the isolated trans-azobenzene monolayer (Fig.~\ref{fig:transazo}, top panel) the first peak corresponds to the first allowed intramolecular $\pi-\pi^*$ transition. 
A dipole-forbidden excitation, again with intramolecular character, is present below 2 eV~\cite{fu+17pccp}.
In the spectrum of cis-azobenzene (Fig.~\ref{fig:transazo}, bottom panel) the lowest-energy excitation is not dark, and gives rise to the weak peak at about 2.8 eV. 
Due to the bended geometry of the cis-configuration, the HOMO-LUMO transition, which is forbidden in the trans-phase, becomes optically allowed.
By comparing the two spectra, the TDA seems to have a rather different effect. 
In trans-azobenzene, the first peak is blue-shifted by approximately 0.3 eV when the TDA is applied.
The weaker maxima at higher energies are also shifted but by a smaller amount ($\sim$0.1 eV).
This behavior reflects the trend discussed above for the biphenyl crystal and is in agreement with the result obtained for the isolated azobenzene molecule~\cite{grue+09nl}.
In Ref.~\cite{grue+09nl}, the mechanism was rationalized as follows. 
The $\pi-\pi^*$ transition giving rise to the first peak has a strong anti-resonant component that contributes to the final excitation energy. 
In the TDA, this component is not coupled to its resonant counterpart such that the resulting peak is overestimated in intensity and by a few hundreds meV in energy.
On the other hand, following the same line of reasoning, the bent geometry of cis-azobenzene reduces the effective C-conjugation of the molecule, and hence the $\pi$ character of their orbitals.
As a result, the coupling between resonant and anti-resonant components of the excitation is reduced, and the TDA spectrum is in agreement with the one obtained from the full BSE.

\section{Summary and Conclusions}
In summary, we have presented the implementation of the $\mathbf{q}$-dependent BSE formalism beyond the Tamm-Dancoff approximation in the all-electron full-potential code \texttt{exciting}.
Our state-of-the-art approach generalizes the previous developments~\cite{Gulans2014,Sagmeister2009} by going beyond the optical limit ($\mathbf{q} \rightarrow 0$) and including the coupling between excitations and de-excitations.
After reviewing the underlying theoretical formalism, we have discussed the specific features of the implementation.
With the aid of selected examples we have shown the capabilities of the developed formalism to describe optical and core excitations. 
In the case of LiF, a prototypical insulator, we have reproduced the dynamical structure factor and determined the excitonic band structure.
Our results are in good agreement with available experiments~\cite{abba+08pnas} and previous theoretical works at the same level of theory~\cite{gatt-sott13prb}.
We have also computed the $\mathbf{q}$-dependent core excitations of LiF from the F K-edge, demonstrating that the selection rules holding in the optical limit ($\mathbf{q} \rightarrow 0$) break down at finite momentum transfer.
The effect of the TDA has been discussed for the loss function of bulk silicon, demonstrating that the interplay between excitonic and plasmonic effects can be properly captured only solving the full BSE~\cite{olev-rein01prl}.
We have also examined the optical spectra of selected organic materials such as biphenyl crystal and azobenzene monolayers, the latter in both the \textit{trans} and \textit{cis} phases.
In the case of biphenyl, the TDA gives rise to an overestimation of the excitation energies by a few hundred meV, as discussed also in Ref.~\cite{pusc+13condmat}.
The same behavior is exhibited also by the spectrum of the trans-azobenzene monolayer, consistent with previous results obtained for the isolated molecule~\cite{grue+09nl}.
Conversely, in the case of cis-azobenzene, the absorption spectrum computed within the TDA is almost identical to the one obtained from the full BSE, indicating that the coupling between the resonant and the anti-resonant components of the excitations decreases with the effective reduction of the $\pi$-conjugation network induced by the bent conformation of the molecule.

The BSE developments presented in this work enlarge the applicability and the predictive power of this formalism to scattering spectroscopic techniques, such as EELS, IXS, and its resonant counterpart (RIXS).
The implementation of the BSE provided in the \texttt{exciting} code allows the application of this demanding methodology to complex systems for both optical and core excitations.

\section*{Acknowledgement}
This work was partially funded by the Deutsche Forschungsgemeinschaft (DFG) - Projektnummer 182087777 - SFB 951 and Projektnummer 12489635 - SFB 658. Additional funding was provided by the Leibniz ScienceCampus ''Growth and Fundamentals of Oxides'' (GraFOx).
\section*{Appendix A: Calculation Flowcharts}
The BSE implementation presented in this work is requires the execution of separate tasks, which are executed sequentially and only interact with each other through binary and human-readable files. This way, calculations can be restarted from each completed task. The same tasks are executed for calculations in the optical and core region. The code differentiates between them within tasks if necessary. In the following, the flowcharts for each tasks is presented.
\begin{figure}[h]
\center
  \includegraphics[width=.5\textwidth]{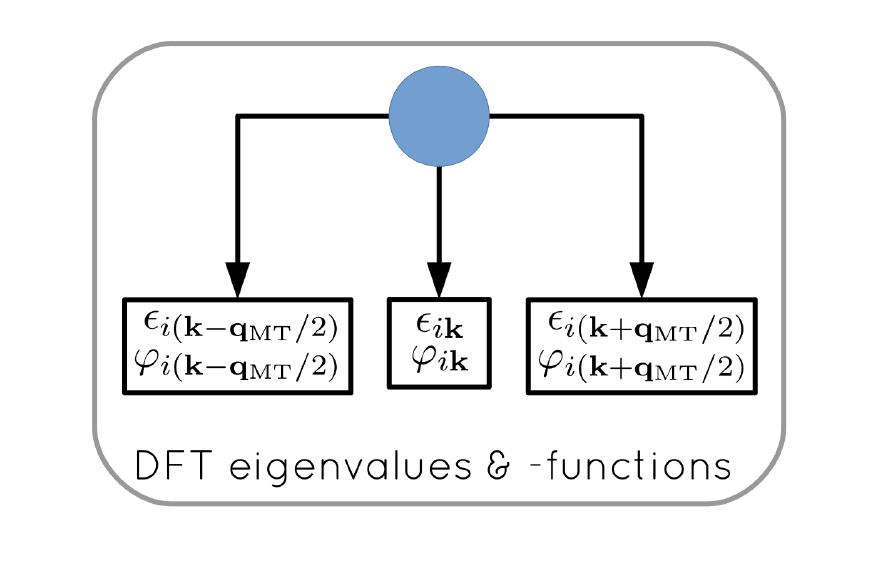}
  \caption{Flowchart for the determination of DFT-eigenvalues and -functions for the BSE calculation. The blue circle indicates a ''loop'', as the DFT calculation is performed thrice for each value $\mathbf{q}_{MT}$ of momentum transfer.}
  \label{fig:task1}
\end{figure}
The first one, shown in Fig.~\ref{fig:task1}, comprises the calculation of the DFT-eigenvalues and -functions needed in the construction of the RPA screening and the BSE matrix elements. For a BSE calculation at finite $\mathbf{q}_{MT}$, DFT calculations with a single self-consistent loop are performed on the $\left(\mathbf{k}\pm \mathbf{q}_{MT}\right)$- and $\mathbf{k}$-grids.
\begin{figure}[h]
\center
  \includegraphics[width=.5\textwidth]{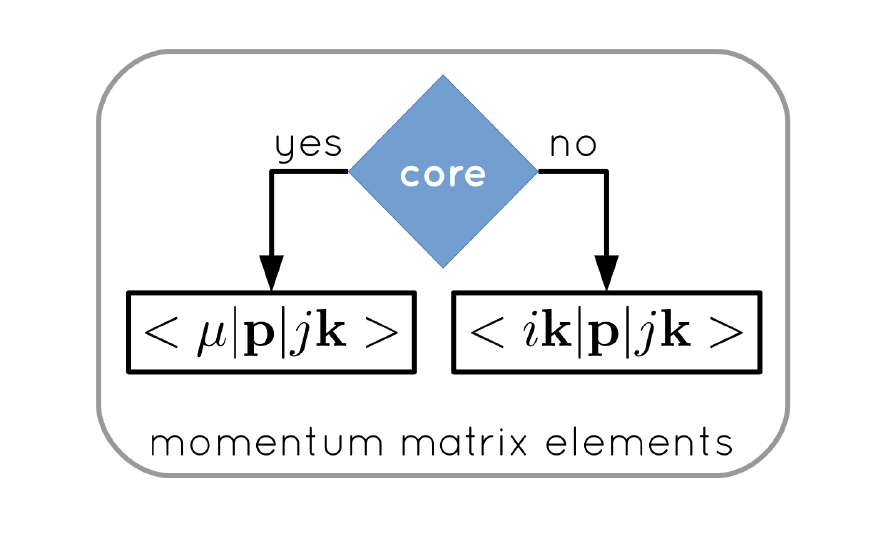}
  \caption{Flowchart for the calculation of momentum matrix elements. The blue triangle inidcates the if-condition, where the code checks whether a core-level calculation shall be performed.}
  \label{fig:task2}
\end{figure}
The second task, shown in Fig.~\ref{fig:task2}, involves the calculation of momentum matrix elements as described in Section IV.A. Depending on whether the BSE calculation in the optical or core region is performed, the valence-conduction or core-conduction momentum matrix elements are calculated and written to file. 
\begin{figure}[h]
\center
  \includegraphics[width=.5\textwidth]{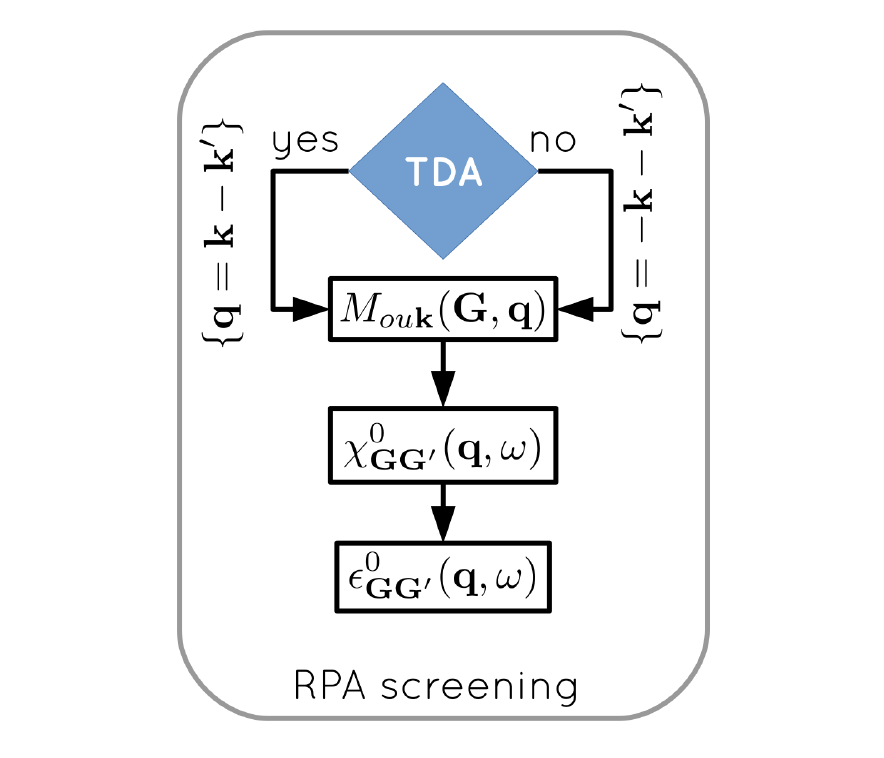}
  \caption{Flowchart for the calculation of the RPA dielectric function.}
  \label{fig:task3}
\end{figure}
In the third task, the RPA screening is calculated. The corresponding flowchart is shown in Fig.~\ref{fig:task3}. Within the TDA, the RPA dielectric function is required on the $\{\mathbf{q}=\mathbf{k}-\mathbf{k}'\}$ set, for calculations beyond the TDA, it additionally has to be calculated for the set  $\{\mathbf{q}=-\mathbf{k}-\mathbf{k}'\}$. For each $\mathbf{q}$, plane-wave matrix elements are calculated, which are then used to determine the independent-particle susceptiblity, and finally the RPA dielectric function. Special care is required in the limit $\mathbf{q} \rightarrow 0$, where the dielectric tensor diverges. The treatment of the divergent terms is discussed in Ref.~\cite{Sagmeister2009}. 
Finally, the dielectric function $\varepsilon_{\mathbf{G}\mathbf{G}'}(\mathbf{q})$ is written to file.
\begin{figure}[h!]
\center
  \includegraphics{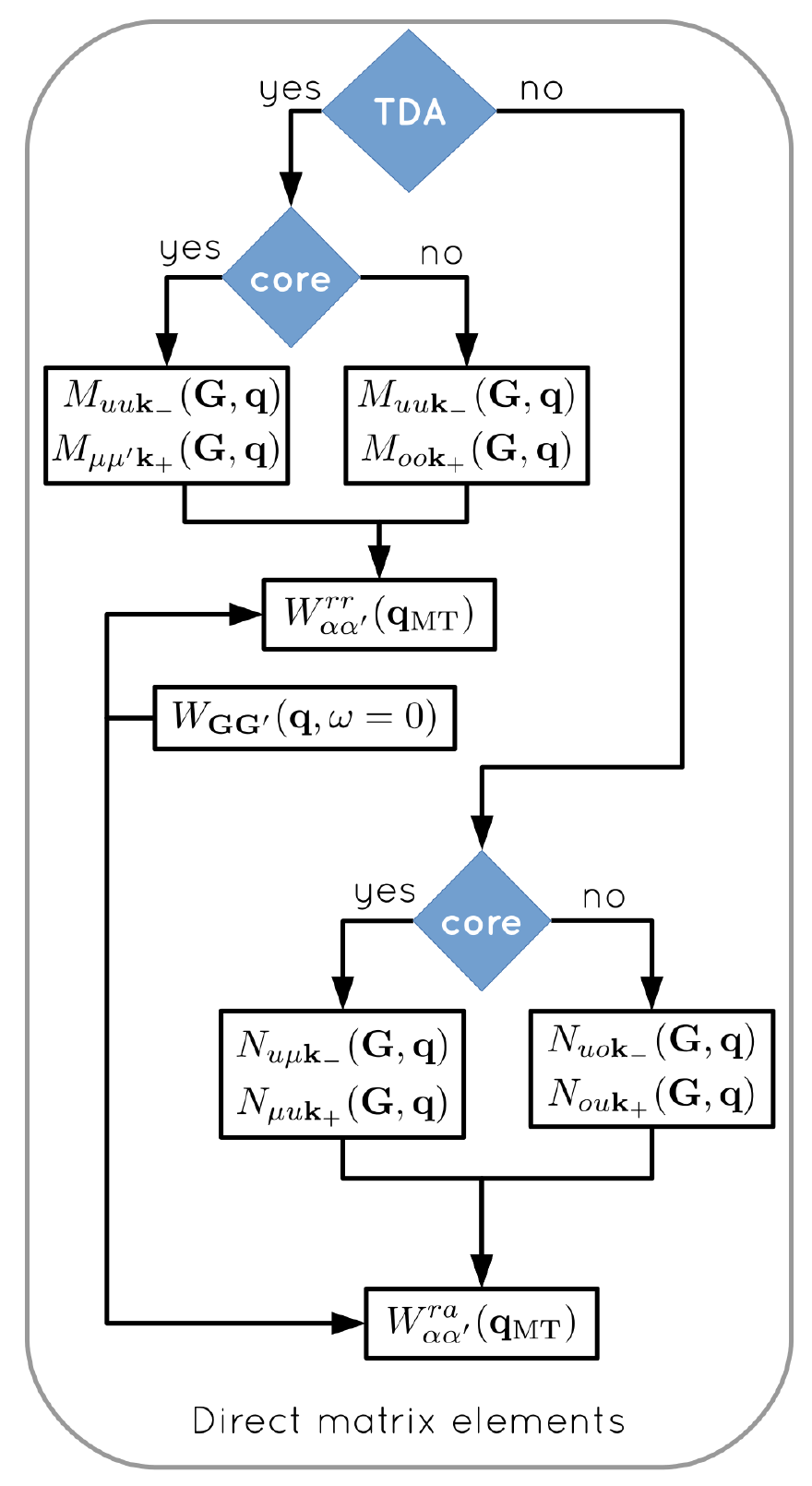}
  \caption{Flowchart for the calculation of the matrix elements of the direct interaction.}
  \label{fig:task4}
\end{figure}
The flowchart for the matrix elements of the direct interaction (Fig.~\ref{fig:task4}) is different depending on whether the TDA is employed and whether core or valence states are the initial states of the transitions. Within the TDA, only the resonant-resonant matrix element $W^{rr}_{\alpha, \alpha'}$ have to be calculated, whereas in a full calculation, resonant-antiresonant matrix elements $W^{ra}_{\alpha, \alpha'}$ have to be determined as well. The screened Coulomb interaction $W_{\mathbf{G}\mathbf{G}'}(\mathbf{q})$ is obtained from the RPA dielectric function calculated according to Fig.~\ref{fig:task3}.
\begin{figure}[h]
\center
  \includegraphics[width=.5\textwidth]{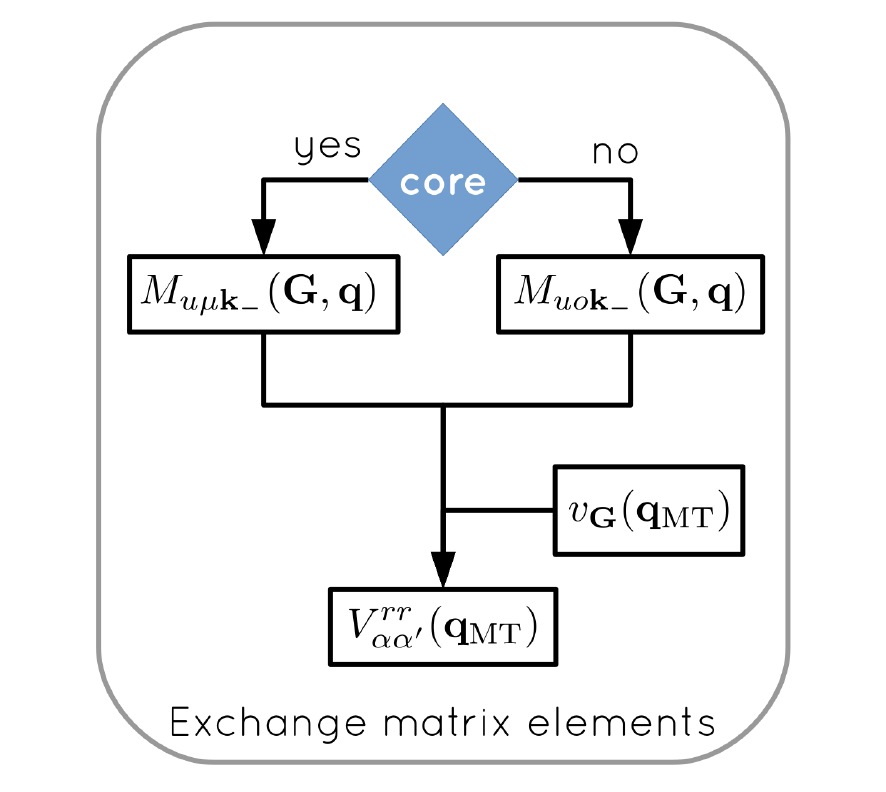}
  \caption{Flowchart for the calculation of the matrix elements of the exchange interaction.}
  \label{fig:task5}
\end{figure}
Figure~\ref{fig:task5} displays the flowchart for the calculation of exchange matrix elements. The calculation differs for optical and core calculations, but is independent of the TDA. 
\begin{figure}[h]
\center
  \includegraphics[width=.5\textwidth]{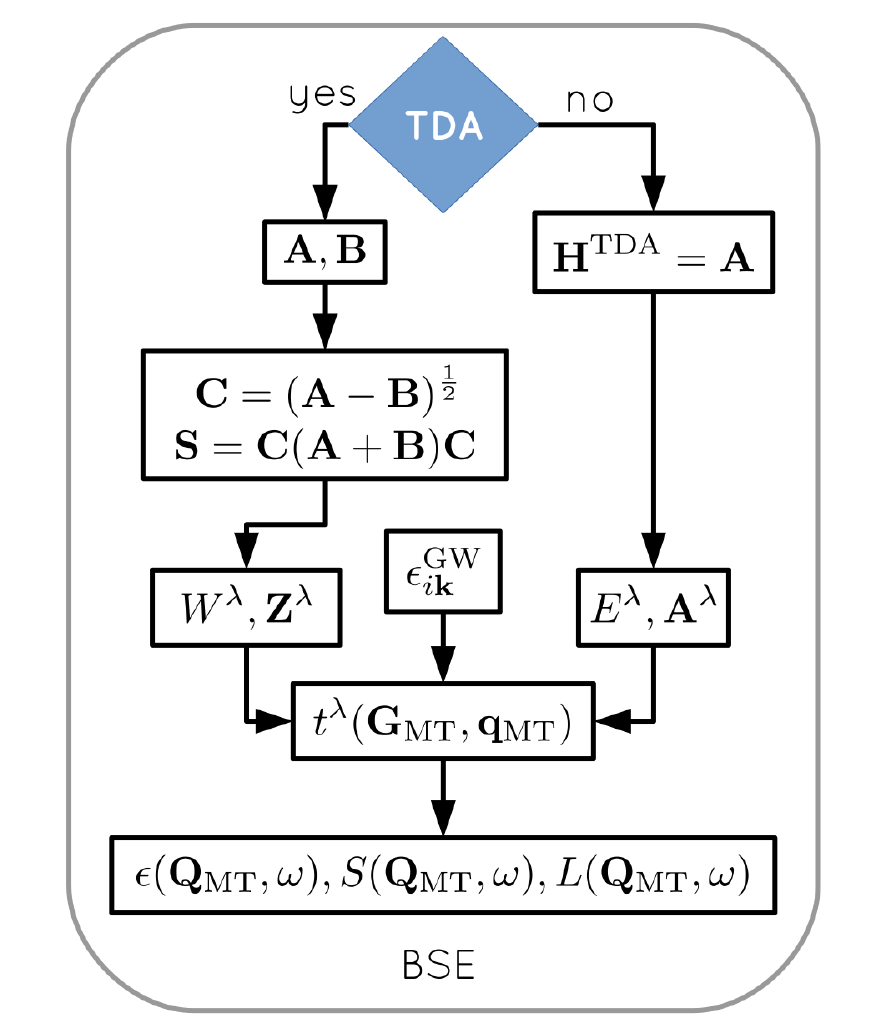}
  \caption{Flowchart for the diagonalization of the BSE Hamiltonian and the construction of the dielectric function.}
  \label{fig:task6}
\end{figure}
In the final task of the BSE implementation (see Fig.~\ref{fig:task5}), the BSE Hamiltonian is constructed and diagonalized. From the eigenstates of the BSE Hamiltonian, the dielectric function, loss function, and dynamical structure factor are calculated.
\section*{Appendix B: Input parameters for BSE calculations in \texttt{exciting}}
\begin{figure}[t]
\center
  \includegraphics[width=.45\textwidth]{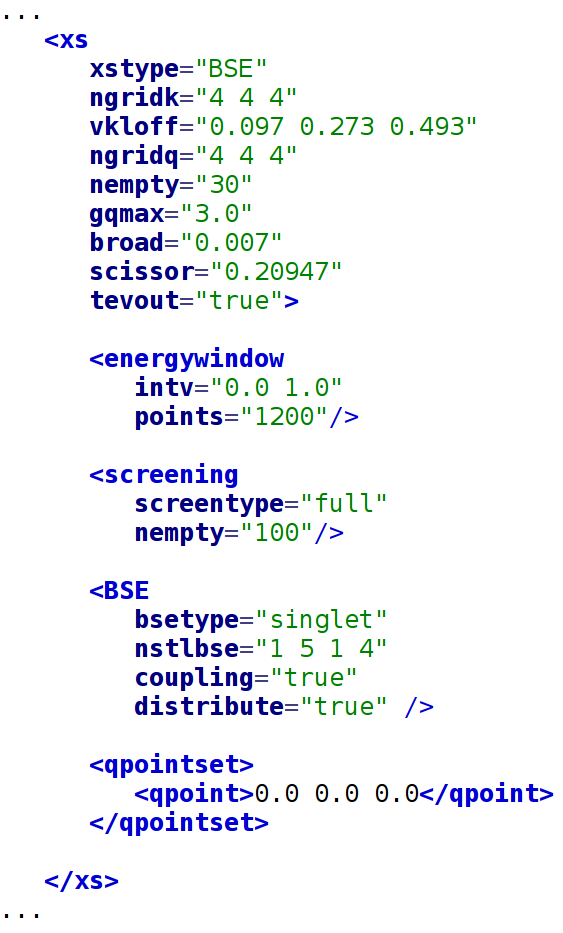}
  \caption{Input XML file for an optical BSE calculation without momentum transfer.}
  \label{fig:input}
\end{figure}

The parameters for calculations with the \texttt{exciting} code are provided through an input file written in the extensible markup language (XML). An overview of the elements and attributes in this input file is provided in Ref.~\cite{Gulans2014}. Full reference can be found in Ref.~\cite{exciting:ref}. Here, we discuss only the input parameters that govern the BSE calculation. We start by considering the input file of LiF at zero momentum transfer (Fig.~\ref{fig:input}). The BSE calculation is triggerd by the presence of the element \texttt{xs}, which includes all the attributes related to excited-state runs.

As attributes of the element \texttt{xs}, we find all the parameters that determine the numerical accuracy of the BSE calculation: \texttt{ngridk}, \texttt{ngridq}, and \texttt{vkloff} define the $\mathbf{k}$- and $\mathbf{q}$-grids and the offset $\mathbf{v}_{off}$. The attribute \texttt{broad} defines the full-width at half maximum of the Lorentzian broadening employed in the calculation, while \texttt{gqmax} defines the plane-wave cut-off for the expansion of matrix-elements and potentials, where only reciprocal lattice vectors $\mathbf{G}$ that fulfill $|\mathbf{G}+\mathbf{q}|\le |\mathbf{G}+\mathbf{q}|_{\mathrm{max}}$ are included. The attribute \texttt{scissor} defines the energy of the scissors operator, which can be applied to the Kohn-Sham energies to open the bandgap.

The element \texttt{xs} contains four required subelements: \texttt{energywindow} defines the energy grid for which the dielectric function, loss function, and dynamical structure factor are calculated; \texttt{screening} determines the parameters for the RPA calculation to obtain the screened Coulomb potential of Eq.~\ref{eqn:wft}; \texttt{BSE} defines the parameter for the actual BSE calculation, and finally \texttt{qpointset}, where values for the momentum transfer $\mathbf{q}$ have to be defined.
\begin{table}
\begin{center}
\begin{tabular}{ |c c c| }
  \hline
  Type of calculation & $\gamma_x$ & $\gamma_c$ \\
  \hline
  \texttt{singlet} & 1 & 1 \\
  \texttt{triplet} & 0 & 1 \\
  \texttt{RPA}     & 1 & 0 \\
  \texttt{IP}      & 0 & 0 \\
  \hline
\end{tabular}
\caption{Possible options for BSE calculations defined by the attribute \texttt{bsetype}}
\label{tab:approx}
\end{center}
\end{table}

The attributes in the element \texttt{BSE} define the numerical parameter for the construction of the BSE Hamiltonian of Eqs.~\ref{eqn:diagonalblock} and \ref{eqn:couplingblock} and of its diagonalization method. The attribute \texttt{bsetype} defines the different types of calculation, depending on the values of $\gamma_x$ and $\gamma_c$ in Eqs.~\ref{eqn:diagonalblock} and \ref{eqn:couplingblock} (see Table~\ref{tab:approx}). The attribute \texttt{nstlbse} contains four integer numbers, which define the transition space, namely the range of occupied (first two numbers) and unoccupied states (last two numbers) included in the BSE Hamiltonian. In the example of LiF shown in Fig.~\ref{fig:input}, the first 5 occupied states counting from the lowest valence one, and the first 4 unoccupied ones are considered. The boolean attribute \texttt{coupling} defines whether the Tamm-Dancoff approximation is used or not, \textit{i.e.} \texttt{coupling}=\texttt{False} triggers a full BSE calculation without the TDA. The attribute \texttt{distribute} defines whether the BSE Hamiltonian is diagonalized with the distributed ScaLapack solver \cite{scalapack} or the serial LaPack solver \cite{lapack}.

\begin{figure}[t]
\center
  \includegraphics[width=.45\textwidth]{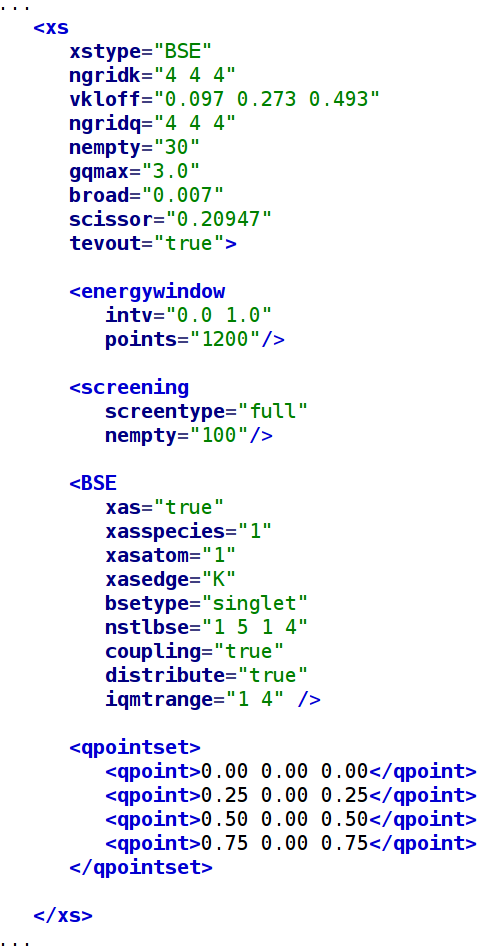}
  \caption{Input XML file for a core-level BSE calculation for different values of momentum transfer along the $\Gamma-X$ direction.}
  \label{fig:input_core}
\end{figure}

For core-level calculations, additional attributes have to be provided to the \texttt{BSE} element. The attributes \texttt{xasspecies}, \texttt{xasatom}, and \texttt{xasedge} specify the species, atom, and edge that is excited, while \texttt{nstlxas} specifies the range of unoccupied states in the BSE Hamiltonian. An example is shown in Fig.~\ref{fig:input_core}.

For calculations at different finite values of the momentum transfer, the $\mathbf{q}$-vectors are defined in the element \texttt{qpointset}, where each vector is provided as a subelement \texttt{qpoint} in units of reciprocal lattice vectors. The range of $\mathbf\mathbf{q}$-vectors included in the calculation is specified in the attribute \texttt{iqmtrange} of the element \texttt{BSE}. An example input is shown in Fig.~\ref{fig:input_core}.
%
\clearpage
\bibliographystyle{iopart-num}
\bibliography{bibliography}

\end{document}